\documentclass[12pt,english,preprint]{article}
\usepackage[T1]{fontenc}
\usepackage[latin9]{inputenc}
\usepackage[a4paper]{geometry}
\usepackage{float}
\usepackage{amsthm}
\usepackage{amsmath}
\usepackage{amssymb}
\usepackage{wasysym}
\usepackage{graphicx}
\usepackage{setspace}
\usepackage{esint}
\makeatletter

\newcommand*\LyXThinSpace{\,\hspace{0pt}}
\providecommand{\tabularnewline}{\\}

\makeatother

\usepackage{babel}
\begin{document}

\title{Gravitational shock wave inside a steadily-accreting spherical charged
black hole}

\author{Ehud Eilon}

\maketitle
\noindent \begin{center}
\emph{Department of Physics, }
\par\end{center}

\noindent \begin{center}
\emph{Technion - Israel Institute of Technology, }
\par\end{center}

\noindent \begin{center}
\emph{Haifa 3200003, Israel}
\par\end{center}
\begin{abstract}
We numerically investigate the interior of a four-dimensional, spherically
symmetric charged black hole accreting neutral null fluid. Previous
study by Marolf and Ori suggested that late infalling observers encounter
an effective shock wave as they approach the outgoing portion of the
inner horizon. Non-linear perturbations could generate an effective
gravitational shock wave, which manifests as a drop of the area coordinate
$r$ from inner horizon value $r_{-}$ towards zero in an extremely
short proper time duration of the infalling observer. We consider
three different scenarios: a) A charged black hole accreting a single
(ingoing) null fluid; b) a charged black hole perturbed by two null
fluids, ingoing and outgoing; c) a charged black hole perturbed by
an ingoing null fluid and a self-gravitating scalar field. While we
do not observe any evidence for a gravitational shock in the first
case, we detect the shock in the other two, using ingoing timelike
and null geodesics. The shock width $\Delta\tau$ decreases rapidly
with a fairly good match to a new, generalized exponential law, $\Delta\tau\sim e^{-\intop\kappa_{-}(\widetilde{V}_{f})d\widetilde{V}_{f}}$,
where $\widetilde{V}_{f}$ is a specific timing parameter for the
ingoing timelike geodesics and $\kappa_{-}(\widetilde{V}_{f})$ is
a generalized (Reissner-Nordström like) surface gravity of the charged
black hole at the inner horizon. We also gain new insight into the
inner (classical) structure of a charged black hole perturbed by two
null fluids, including strong evidence for the existence of a spacelike
$r=0$ singularity. We use a finite-difference numerical code with
double-null coordinates combined with an adaptive gauge method in
order to solve the field equations from the region outside the black
hole down to the vicinity of the $r=0$ singularity.

\newpage{}
\end{abstract}

\section{INTRODUCTION \label{sec:introduction}}

\subsection{Background}

The study of the inner structure of classical black holes (BHs) has
been an enduring research field in the last couple of decades. Two
classes of models were of a particular interest; perturbations on
a charged BH background \cite{1-Penrose-PRN-Analytical,2-Simpson-Penrose-PRN-Numerical,4-Novikov-Gursel-PRN,3-Hiscock-PRN-Analytical,5-Chand-Hartle-PRN-Analytical,6-Poisson-Israel-Letter,7-Poisson-Israel-mass-function,8-Ori-DNF,9-Gnedin,10-Brady-Smith,Burko-added1,11-Burko-Internal-structure,12-Hod-Piran,13-Burko-EOMs,14-Burko-spacelike,Burko-added2,Burko-added3,15-Dafermos-stability}
and perturbations on a spinning BH background \cite{16-Ori-spinning,17-Brady-Morinsk-Spinning,18-Ori-spinning-perturbation,19-Ori-Spinning,Hamilton-Added1,Hamilton-Added2,Hamilton-Added3,20-Burko-CScalars-Kerr}.
The research considered the effects of perturbations on the BH geometry
and the differences from the corresponding non-perturbed geometries,
the Reissner-Nordström (RN) geometry in the charged case and the Kerr
geometry in the spinning case. Both classes of models share a similar
horizon structure (of an (outer) event horizon (EH) and an inner horizon
(IH)); while the perturbed spinning class is believed to be closely
related to realistic astrophysical BHs, the perturbed charged class
(usually) offers a simpler analysis due to spherical symmetry. RN
geometry has a well known $r=0$ timelike singularity; Kerr geometry
has a timelike $r=0$ ring singularity. The study of the perturbed
geometries focused specifically on the development of additional singularities
inside the BHs. 

The IH of RN and Kerr geometries also operates as Cauchy Horizon (CH),
a null hypersurface that marks the boundary of physical predictability.
\footnote{The overlap between the IH and CH is not necessarily full; the IH
has two arms, distinguished by their null direction (ingoing ($u$)
or outgoing $(v$)). In a typical gravitational collapse scenario,
only one arm of the IH is CH (the ingoing one).%
} Penrose pointed out that this hypersurface is a locus of infinite
blueshift in both geometries; \cite{1-Penrose-PRN-Analytical} he
predicted it should develop curvature singularity in the presence
of perturbations. Hiscock confirmed this prediction with analytical
analysis of an ingoing null fluid perturbation. \cite{3-Hiscock-PRN-Analytical}
Hiscock used the Reissner-Nordström-Vaidya (RNV) model, \cite{1951-RNV-Vaidya}
representing a neutral null fluid --- a stream of massless particles
--- flowing on a charged black hole background. This model has two
variants, ingoing and outgoing, distinguished by the null direction
of the fluid; Hiscock used the ingoing variant. He discovered that
a nonscalar curvature singularity develops at the ingoing section
of the IH (which is CH in this scenario). Some time later, Poisson
and Israel have diagnosed the development of a singularity at the
CH of a different model, the mass inflation model; \cite{6-Poisson-Israel-Letter,7-Poisson-Israel-mass-function}
this model includes two null fluids, ingoing and outgoing, flowing
on a charged BH background. In this case, however, the singularity
is scalar and there is a divergence of the mass function at the vicinity
of the singular CH. Despite of this, Ori has later proved that this
null singularity is deformationally weak \cite{8-Ori-DNF} in the
Tipler sense (\cite{Tipler-What-is-Weak}, see also Ref. \cite{Ori-What-is-Weak});
the metric tensor components approach a finite value on the CH and
an infalling observer only experiences finite tidal distortion at
the crossing of the CH. During the 1990's, numerical investigations
of self-gravitating scalar field perturbations on a charged background
\cite{10-Brady-Smith,11-Burko-Internal-structure} suggested that
in this case the singular CH is a subject to a process of ``focusing'';
the area coordinate $r$ monotonically decreases along CH up to the
point where it vanishes and the singularity becomes spacelike. The
analysis of perturbations on a spinning background revealed a similar
general picture; \cite{16-Ori-spinning,17-Brady-Morinsk-Spinning,18-Ori-spinning-perturbation,19-Ori-Spinning,20-Burko-CScalars-Kerr}
in this case too, a weak null curvature singularity develops at the
CH.

\subsection{Shock wave at the inner horizon}

Marolf and Ori (MO) have recently demonstrated analytically the existence
of additional null (effective) singularity at the outgoing portion
of the IH, an effective shock wave singularity. \cite{Marolf-Ori_Shockwave}
An infalling observer experiences a finite (effective) jump in the
values of various perturbation fields across the IH at late times.
This change actually occurs on a finite proper time duration $\Delta\tau$;
however, this duration decreases exponentially with infall time and
becomes unresolvable for an observer of given sophistication. The
observer may experience a metric discontinuity if the perturbation
is nonlinear; this metric discontinuity, or gravitational shock, manifests
as an (effective) sheer drop in the value of the area coordinate $r$
(where $g_{\theta\theta}=r^{2}$). 

MO have considered specifically a wide variety of test perturbations
(scalar field, electromagnetic and gravitational perturbations) on
RN and Kerr backgrounds and non-linear scalar field perturbation on
a spherical charged background. They estimated that their arguments
should be relevant to non-linear perturbations on a spinning BH background
as well but did not include a full analysis of such case. Their scenarios
were asymptotically flat and asymptotically static; they relied upon
known properties of RN and Kerr geometries. MO argued that the shock
wave singularity is ``stronger'' and more violent than the CH singularity;
an infalling observer experiences integrated deformation across the
IH which does not decrease with infall time, but remains fixed and
of order unity.

More recently, Eilon and Ori (EO) have confirmed numerically the existence
of the shock wave. \cite{Eilon-Ori_Shockwave} EO considered two different
scenarios, the evolution of a test scalar-field on RN background and
the evolution of a self-gravitating scalar-field on a dynamical charged
BH background. They have demonstrated the existence of the shock in
the scalar field $\Phi(\tau)$ in both cases, and in $r(\tau)$ in
the self-gravitating case. EO also confirmed MO's prediction about
the exponential decrease in $\Delta\tau$ (the exponential sharpening
rate) of the shock; they have defined characteristic $\Delta\tau$
widths for both $\Phi(\tau)$ and $r(\tau)$ and showed that they
decrease exponentially with infall time. Although most of their analysis
was based on timelike geodesics, EO were the first to exhibit the
shock on null geodesics; they displayed the gravitational shock in
$r(\lambda)$ (where $\lambda$ is the affine parameter) on a three-dimensional
graph using a dense set of null geodesic. The geodesics' sheer drop
in $r$ at the IH created a vertical wall-like structure; they argued
that this may be the clearest visual presentation of the shock. 

Fig. \ref{fig:Penrose-SGSF} illustrates the gravitational shock scenario
considered by EO through the relevant Penrose diagram. The inner (classical)
structure of the BH in this case is well known (with the possible
exception of the shock wave), and includes a strong spacelike $r=0$
singularity and a weak null singularity at the CH. The shock is located
at the outgoing IH of the BH at late times (the solid green line).
The full blue curve and the dashed blue line represent a late timelike
geodesic and a late null geodesic, accordingly; both cross the IH
and reach the spacelike $r=0$ singularity. Due to the gravitational
shock, the journey from the IH ($r=r_{-})$ to $r=0$ takes an extremely
short proper time duration $\Delta\tau$ (or an affine parameter interval
$\Delta\lambda$). %
\footnote{The affine parameter $\lambda$ has normalization freedom; however,
for any fixed choice of normalization constant $\Delta\lambda$ still
decreases rapidly with infall time, as the ``vertical wall'' picture
of EO demonstrated.%
} Fig. \ref{fig:Penrose-SGSF} describes the case of a self gravitating
scalar field perturbation on an eternal RN background. However, EO
shock results could be also attributed to the case of a spherical
charged collapse; they could describe the dynamics outside a collapsing
charged shell or star. %
\footnote{A small region inside the BH at the vicinity of RN original timelike
$r=0$ singularity (in the left border of Fig. \ref{fig:Penrose-SGSF})
should be omitted from the numerical results in order to make this
interpratation valid; however, the shock is not influenced by this
omission. %
} The shock wave phenomenum, as argued by MO and EO, is a general phenomenum
of perturbed charged or spinning BHs; it is not limited to the eternal
RN (or Kerr) scenarios. 

\begin{center}
\begin{figure}[H]
\noindent \begin{centering}
\includegraphics[scale=0.5]{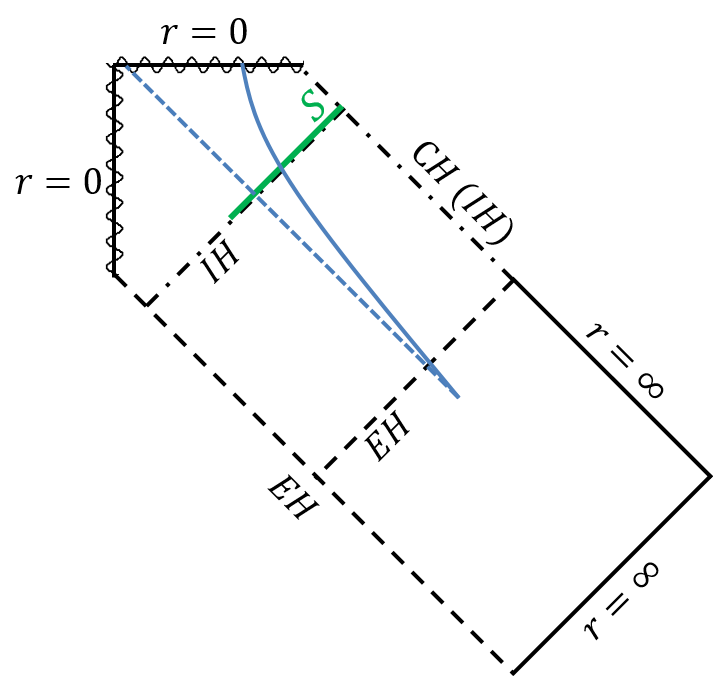}
\par\end{centering}

\protect\caption{\label{fig:Penrose-SGSF} Penrose diagram of a charged BH perturbed
by a self gravitating scalar field. Solid black diagonal lines denote
null infinity; dashed black diagonal lines denote the event horizon
(EH). Dashed-dotted line denotes the inner horizon (IH); the ingoing
IH functions as a Cauchy horizon (CH). Wavy lines denote the timelike
$r=0$ singularity of the initial RN spacetime and the spacelike $r=0$
singularity of the perturbed spacetime. An effective gravitational
shock wave develops along the solid green line denoted ``S'' (at
the outgoing IH). The full blue line represents a typical timelike
geodesic that crosses the shock; the dashed blue line is a typical
null geodesic. The shock manifests as a sharp drop in the values of
the area coordinate $r$ along both geodesics, from $r_{-}$ to zero,
in an extremely short proper time duration $\Delta\tau$ (timelike
geodesic) or an affine parameter interval $\Delta\lambda$ (null geodesic).}
\end{figure}

\par\end{center}

EO considered a single type of perturbation, a neutral massless scalar-field.
They also considered a relatively simple scenario, asymptotically
flat BH accreting a single scalar field pulse that decays at late
times. However, as Hamilton and Avelino has pointed out, \cite{Hamilton-Avelino}
realistic astrophysical BHs steadily accrete dust and cosmic microwave
background (CMB) photons on cosmological timescales. We wish to extend
EO research to a more realistic accretion scenario, and to a different
type of perturbation, a neutral null-fluid, which could represent
(up to some extent) CMB photons. Due to scope limitations, we focus
our analysis on the gravitational shock, the shock in the area coordinate
$r$. We consider in the current paper three different physical scenarios;
all three include long term accretion of a null-fluid stream by a
charged BH. In the first scenario, the ingoing null-fluid stream is
the only perturbation; we do not detect a gravitational shock in this
case. In the second and third scenarios we add an outgoing null-fluid
pulse and a self-gravitating scalar-field pulse (accordingly) to the
ingoing null-fluid stream. We detect a gravitational shock wave in
both scenarios, and uncover different behaviour of the shock due to
the accretion of the null-fluid stream. The sharpening rate of the
shock differs from EO (and MO) result --- our shock sharpens more
rapidly, with a fairly good match to a generalized exponential sharpening
rate law. MO and EO used (mostly) timelike geodesics in their shock
analysis, though EO had some results taken from null geodesic. We
employ both timelike and null geodesics, alternately, throughout our
entire analysis. In the course of our shock exploration we uncover
two interesting results on the inner (classical) structure of a charged
BH perturbed by two null fluids --- a strong evidence for the existence
of a spacelike $r=0$ singularity and a possible, uncertain indication
to the presence of a null, non-naked, $r=0$ singularity. 

The motivation for our third scenario, a mixed null-fluid/scalar field
perturbation, requires a further explanation. The dynamics of scalar
field perturbation are different from those of null fluid perturbation
and more interesting in the sense that they are more similar to those
of realistic gravitational or electromagnetic perturbations. EO have
already demonstrated the presence of a shock in the case of self gravitating
scalar field perturbation; here we are interested in observing the
effect of the null fluid stream on a previously generated shock. This
scenario also allows a better comparison of our shock results with
EO results than the two null fluids case, as the mixed scenario is
basically EO's scenario with the addition of an ingoing null fluid
stream. Lastly, the mixed scenario is a ``toy model'' for an astrophysical
BH accreting both dust (simulated by the scalar field) and radiation
(simulated by the null fluid). 

The paper has the following structure: The problem is formulated in
terms of unknown functions, field equations, and initial data setup
in Sec. \ref{sec:Field-Equations}. The description in this section
is mostly general and applies for all three scenarios, which are distinguished
from each other by initial data setup. We sketch this separation schematically
in Sec. \ref{sec:Field-Equations}; the details are described in Secs.
\ref{sec:RNV}-\ref{sec:mixed}. The numerical algorithm is discussed
in Sec. \ref{sec:Numerical-Algorithm}; we describe in this section
the numerical solver of the field equations and outline additional
important calculations and the presentation of numerical results.
Since the field equations and the numerical algorithm were discussed
extensively in a previous paper \cite{Eilon-Ori_AdaptiveGauge} and
are quite similar to those used by EO, we only describe them briefly
here, focusing on the differences from EO setup. We then analyze the
case of a single ingoing null-fluid stream on a charged background
in Sec. \ref{sec:RNV}. In particular, we exhibit the lack of evidence
for gravitational shock. The case of two null fluids (ingoing and
outgoing) on a charged background is analyzed in Sec. \ref{sec:DNF},
where we demonstrate the existence of a gravitational shock wave and
analyze the rate of shock sharpening. We do the same for the third
case, an ingoing null-fluid stream with a self-gravitating scalar-field
pulse on a charged background, in Sec. \ref{sec:mixed}. We summarize
and discuss our results in Sec. \ref{sec:discussion}.

\section{FIELD EQUATIONS \label{sec:Field-Equations}}

We consider in this paper three different physical scenarios: (i)
a preexisting RN BH accreting a single (ingoing) null-fluid; (ii)
two null fluids, ingoing and outgoing, flowing on a charged BH background;
(iii) a preexisting RN BH accreting a a single (ingoing) self-gravitating
scalar-field pulse and a single (ingoing) null-fluid. We investigate
these cases using the same set of field equations solved by the same
numerical algorithm; they are distinguished by initial conditions
choice (see section \ref{sub:Char-IC}).

The null fluid is neutral and minimally coupled. The scalar field
is the same self-gravitating scalar field used by EO; it is uncharged,
massless and minimally coupled, satisfying the massless Klein-Gordon
equation $\Square\Phi=0$. In all three cases the initial RN geometry
has mass $M_{0}$ and charge $Q$. The line element in double-null
coordinates $(u,v,\theta,\varphi)$ is

\begin{equation}
ds^{2}=-e^{\sigma(u,v)}dudv+r(u,v)^{2}d\Omega^{2}\,,\label{eq:line-element}
\end{equation}

\noindent where $d\Omega^{2}\equiv d\theta^{2}+\sin^{2}\theta\, d\varphi^{2}$.
In principle, our three unknown functions are the metric functions
$\sigma(u,v)$ and $r(u,v)$ and the scalar field $\Phi(u,v)$, although
in the single null fluid and two null fluids cases the scalar field
is trivially solved ($\Phi(u,v)=0$). 

The field equations are given by $G_{\mu\nu}=8\pi(T_{\mu\nu}^{\Phi}+T_{\mu\nu}^{Q}+T_{\mu\nu}^{NF})$,
where $T_{\mu\nu}^{\Phi}$ and $T_{\mu\nu}^{Q}$ are the energy-momentum
tensors of the scalar and electromagnetic fields; $T_{\mu\nu}^{NF}$
is the energy-momentum tensor of the null fluid and satisfies (see
e.g. Ref. \cite{7-Poisson-Israel-mass-function})

\begin{equation}
T_{\mu\nu}^{NF}=\rho_{in}k_{\mu}k_{\nu}+\rho_{out}l_{\mu}l_{\nu}\:,\label{eq:null-fluid-emt}
\end{equation}

\noindent where $\rho_{in},\rho_{out}$ are constants, $k_{\mu}$
a radial null vector pointing inward and $l_{\mu}$ a radial null
vector pointing outward. This tensor has only two nonvanishing components
in double null coordinates, $T_{vv}^{NF}$ and $T_{uu}^{NF}$. $T_{vv}^{NF}$
corresponds to the first term in Eq. (\ref{eq:null-fluid-emt}), the
ingoing null fluid; $T_{uu}^{NF}$ corresponds to the second term,
the outgoing null fluid. We henceforth define the null fluid by these
components (rather than $\rho_{in},\rho_{out},k_{\mu}$ and $l_{\mu})$
for the sake of simplicity. Overall, our field equations consist of
three evolution equations, 

\noindent 
\begin{equation}
r,_{uv}=-\frac{r,_{u}r,_{v}}{r}-\frac{e^{\sigma}}{4r}(1-\frac{Q^{2}}{r^{2}})\,,\label{eq:  r_evolutio}
\end{equation}
\begin{equation}
\sigma,_{uv}=\frac{2r,_{u}r,_{v}}{r^{2}}+\frac{e^{\sigma}}{2r^{2}}(1-\frac{2Q^{2}}{r^{2}})-2\Phi,_{u}\Phi_{,v}\,\,,\label{eq:  sigma_evolutio}
\end{equation}

\noindent 
\begin{equation}
\Phi,_{uv}=-\frac{1}{r}(r,_{u}\Phi,_{v}+r,_{v}\Phi,_{u})\,,\label{eq:  phi_evolutio}
\end{equation}

\noindent and two constraint equations, 
\begin{equation}
r,_{uu}-r,_{u}\sigma,_{u}+r(\Phi,_{u})^{2}+4\pi rT_{uu}^{NF}=0\,,\label{eq:  ruu}
\end{equation}
\begin{equation}
r,_{vv}-r,_{v}\sigma,_{v}+r(\Phi,_{v})^{2}+4\pi rT_{vv}^{NF}=0\,.\label{eq:  rvv}
\end{equation}

\noindent The derivation of the field equations is fairly standard
and, with the exclusion of the null fluids terms, has been discussed
extensively in a previous paper \cite{Eilon-Ori_AdaptiveGauge} (in
Sec. II and the appendix) and in other works. \cite{13-Burko-EOMs}
The addition of the null fluid to the model is trivial, however; it
contributes a single term for each of the constraint equations (\ref{eq:  ruu})
and (\ref{eq:  rvv}) while the evolution equations remain unchanged.
The evolution and constraint equations are consistent; the constraint
equations need only be imposed at the initial rays. Hence $T_{uu}^{NF}$
and $T_{vv}^{NF}$ do not have ``evolution equations''; they are
calculated on every point of the grid from the constraint equations.

\subsection{Characteristic initial conditions\label{sub:Char-IC}}

The characteristic initial hypersurface includes two null rays, $u=u_{0}$
and $v=v_{0}$. We choose four functions on each initial ray, which
correspond to two initial conditions for the unknowns $\sigma$ and
$\Phi$ and two initial functions for the energy-momentum components
$T_{uu}^{NF}$ and $T_{vv}^{NF}$. The function $r$ is determined
by a choice of a single parameter --- $r(u_{0},v_{0})\equiv r_{0}$
--- and numerical solution of the constraint equations --- Eq. (\ref{eq:  ruu})
at $v=v_{0}$ and Eq. (\ref{eq:  rvv}) at $u=u_{0}$.

The line element (\ref{eq:line-element}) implies that initial conditions
choice for $\sigma$ is equivalent to a gauge choice for the null
coordinates $u$ and $v$. A gauge transformation $v\rightarrow v'(v),u\rightarrow u'(u)$
does not change $r$ or $\Phi$, but it does change $\sigma$, according
to 

\begin{equation}
\sigma\rightarrow\sigma'=\sigma-\ln(\frac{du'}{du})-\ln(\frac{dv'}{dv})\,\,.\label{eq:  sigma_gauge}
\end{equation}

\noindent Our initial conditions choice for $\sigma$ corresponds
to the maximal-$\sigma$ gauge, defined by 

\begin{equation}
\sigma(u_{0},v)=0\,,\,\:\sigma_{max}(u)=0\,,\label{eq: Sigma_max:u+v}
\end{equation}

\noindent where $\sigma_{max}(u)$ is a function which specifies the
maximal value of $\sigma$ on each constant $u$ line (at the range
$v_{0}\leq v\leq v_{max})$. This adaptive gauge addresses and solves
a numerical resolution loss problem, inherent to long-time simulations
in double-null coordinates near the EH (see Ref. \cite{Eilon-Ori_AdaptiveGauge}).
The gauge condition $\sigma_{max}(u)=0$ translates to initial condition
on $\sigma(u,v_{0})$ via extrapolation procedure, explained in Sec.
VII of Ref. \cite{Eilon-Ori_AdaptiveGauge}.

Table \ref{tab:IC-def} outlines the initial functions choice for
our scenarios. We focus here at the fundamental differences between
the scenarios --- which initial functions vanish and which do not
--- and not on the details of the non-vanishing functions, described
later at the relevant sections (\ref{sub:IC-RNV}, \ref{sub:IC-DNF}
and \ref{sub:IC-Mixed}). Note that $\sigma$ and $T_{vv}^{(NF)}$
have the same basic definition in all three cases. 

\noindent 
\begin{table}[H]
\noindent \begin{centering}
\begin{tabular}{|c|c|c|c|}
\hline 
Function\textbackslash{}Scenario & Single null fluid & Two null fluids & Mixed \tabularnewline
\hline 
\hline 
$\sigma$ & $\begin{array}[t]{c}
\sigma(u_{0},v)=0\\
\sigma_{max}(u)=0
\end{array}$ & $\begin{array}[t]{c}
\sigma(u_{0},v)=0\\
\sigma_{max}(u)=0
\end{array}$ & $\begin{array}[t]{c}
\sigma(u_{0},v)=0\\
\sigma_{max}(u)=0
\end{array}$\tabularnewline
\hline 
$\Phi$ & $\begin{array}[t]{c}
\Phi(u_{0},v)=0\\
\Phi(u,v_{0})=0
\end{array}$ & $\begin{array}[t]{c}
\Phi(u_{0},v)=0\\
\Phi(u,v_{0})=0
\end{array}$ & \emph{$\begin{array}[t]{c}
\Phi(u_{0},v)\neq0\\
\Phi(u,v_{0})=0
\end{array}$}\tabularnewline
\hline 
$T_{uu}^{NF}$ & $\begin{array}[t]{c}
T_{uu}^{NF}(u_{0},v)=0\\
T_{uu}^{NF}(u,v_{0})=0
\end{array}$ & $\begin{array}[t]{c}
T_{uu}^{NF}(u_{0},v)=0\\
T_{uu}^{NF}(u,v_{0})\neq0
\end{array}$ & $\begin{array}[t]{c}
T_{uu}^{NF}(u_{0},v)=0\\
T_{uu}^{NF}(u,v_{0})=0
\end{array}$\tabularnewline
\hline 
$T_{vv}^{NF}$ & $\begin{array}[t]{c}
T_{vv}^{NF}(u_{0},v)\neq0\\
T_{vv}^{NF}(u,v_{0})=0
\end{array}$ & $\begin{array}[t]{c}
T_{vv}^{NF}(u_{0},v)\neq0\\
T_{vv}^{NF}(u,v_{0})=0
\end{array}$ & $\begin{array}[t]{c}
T_{vv}^{NF}(u_{0},v)\neq0\\
T_{vv}^{NF}(u,v_{0})=0
\end{array}$\tabularnewline
\hline 
\end{tabular}
\par\end{centering}

\protect\caption{Initial functions choice for the three different physical scenarios
considered in this paper. The table outlines the separation between
the scenarios by their initial functions choice; it is created by
different choices for the energy momentum tensor component of the
outgoing null fluid ($T_{uu}^{NF}$) and the scalar field $\Phi$.
The table also contains initial functions choice for the energy momentum
tensor component of the ingoing null fluid ($T_{vv}^{NF}$) and the
metric function $\sigma$. The condition $\sigma_{max}(u)=0$ refers
to the maximal-$\sigma$ gauge condition; the maximal value of $\sigma$
on each $u=const.$ grid ray is set to be zero. This condition translates
to initial condition on $\sigma(u,v_{0})$ via extrapolation procedure,
explained in Sec. VII of Ref. \cite{Eilon-Ori_AdaptiveGauge}. \label{tab:IC-def}}
\end{table}

\subsection{Black hole mass and surface gravity\label{sub:Mass-function}}

Our shock analysis requires an estimation of the (growing) BH mass
during the simulation and its changing surface gravity at the IH.
We use the mass function $m(u,v)$ introduced in Ref. \cite{7-Poisson-Israel-mass-function},
which in our coordinates translates to

\noindent 
\begin{equation}
m=(1+4e^{-\sigma}r,_{u}r,_{v})r/2+Q^{2}/2r\,.\label{eq:mass_formula}
\end{equation}

We also follow EO ansatz for the event horizon location --- the (first)
$u$ value where $r,_{v}$, evaluated at the final ingoing ray of
the numerical grid $v=v_{max}$, changes its sign from positive to
negative. %
\footnote{This $u$ value typically falls between two grid points in the numerical
simulation. We find the exact $u_{h}$ value via standard interpolation
procedure. Given the two points $(u_{h1},v_{max})$ and $(u_{h2},v_{max})$
and their $r,_{v}$ values $r,_{v}^{h1}$ and $r,_{v}^{h2}$ (where
$r,_{v}^{h1}$ is the last positive value of $r,_{v}$ and $r,_{v}^{h2}$
is the first negative value of $r,_{v}$), we estimate $u_{h}$ as
$u_{h}=\frac{u_{h2}r,_{v}^{h1}-u_{h1}r,_{v}^{h2}}{r,_{v}^{h1}-r,_{v}^{h2}}$
. All the numerical results on the EH are evaluated in the same fashion.%
} We denote this value as $u_{h}$. We define the black hole mass $m_{BH}(v)$
as the value of the mass function along the event horizon,

\[
m_{BH}(v)\equiv m(u_{h},v)\,.
\]

\noindent This is a monotonically increasing function in our simulation.
The values of the EH and IH also changes, according to

\begin{equation}
r_{\pm}(v)=m_{BH}(v)\pm\sqrt{m_{BH}^{2}(v)-Q^{2}}\,,\label{eq:r_plus_minus}
\end{equation}

\noindent and they imply a steady change in the BH surface gravity
at the EH and IH,

\begin{equation}
\kappa_{\pm}(v)=\frac{\sqrt{m_{BH}^{2}(v)-Q^{2}}}{r_{\pm}^{2}(v)}\,.\label{eq:kappa_plus_minus}
\end{equation}

\noindent We specifically denote the values of the BH mass, EH, IH
and surface gravity at last ingoing ray $v=v_{max}$ as $m_{BH,vmax},r_{\pm,vmax}$
and $\kappa_{\pm,vmax}$ accordingly. %
\footnote{\label{fn:Kappa-is-RN-i}Note that we use the exact RN expression
for the surface gravity. While it could be argued to be relevant for
the single null fluid case as well, its relevance for the case of
two null fluids and the mixed case is less clear. Nevertheless, we
find this expression useful in our analysis and it enables comparison
with MO and EO results. %
}

\noindent The functions $r_{\pm}(v)$ and $\kappa_{\pm}(v)$ could
be reexpressed in terms of a different advanced null coordinate $V(v)$
with the appropriate gauge transformation, as we indeed do below.

\section{NUMERICAL ALGORITHM\label{sec:Numerical-Algorithm}}

We solve the field equations on a double-null grid; the grid has fixed
spacings in $u$ and $v$ (denoted $\Delta u,\Delta v$). We choose
$\Delta u=\Delta v=\frac{M_{0}}{N}$, where $M_{0}$ is the initial
BH mass and $N$ is assigned several different values on each run
(usually $80,160,320,640$), in order to confirm numerical convergence.
The numerical solution begins on the initial ray $u=u_{0}$ and progresses
towards the final ray $u=u_{max}$; along each $u=const$ ray the
solution is advanced from $v=v_{0}$ to $v=v_{max}$. %
\footnote{The numerical solution on the initial rays $u=u_{0}$ and $v=v_{0}$
is actually a solution of ODEs, not PDEs; as explained in Sec. \ref{sub:Char-IC},
we solve the constraint equations (\ref{eq:  ruu}) and (\ref{eq:  rvv})
in order to find $r(u,v_{0})$ and $r(u_{0},v)$, accordingly.%
} The evolution equations (\ref{eq:  r_evolutio}-\ref{eq:  phi_evolutio})
are discretized using a standard finite-differences scheme; we also
apply a predictor-corrector scheme with second order accuracy, as
described in detail in Sec. III of Ref. \cite{Eilon-Ori_AdaptiveGauge}.
The numerical convergence of our unknown functions is usually second
order convergence; however, there is a typical decline in performance
at the close vicinity of the $r=0$ singularity ($r\sim0.1$ or less,
with some variations) to first order convergence. This effect is (partially)
caused by numerical (artificial) fluxes generated by the solution
of the constraint equation (\ref{eq:  ruu}) at the close vicinity
of the singularity. In order to avoid these fluxes, we choose $u_{max}$
value such as $r(u,v_{0})$ never falls below $0.1$.

\subsection{Geodesics definitions\label{sub:Geodesics-definitions}}

The previous section described the generation of results on a numerical
double-null grid; however, shock analysis requires consideration of
results on ingoing geodesics. MO's original analysis had considered
timelike geodesics; EO demonstrated that null geodesics could be effective
as well for the exploration of the gravitational shock, although they
mainly focused on timelike geodesics. We consider here alternately
timelike and null geodesics. Since the derivation of both types has
already been described in detail by EO, we give here just a quick
summary of geodesics definitions and behaviour, focusing on the differences
from EO setup. A full description of the derivation could be found
at the appendix of EO.

\subsubsection{Timelike geodesics\label{sub:T-geo-def}}

Timelike geodesics are more ``physical'' than null geodesics in
the sense that they describe the trajectories of actual (test) observers
or probes falling into the BH. In our double-null grid, timelike geodesics
are a series of bent curves $v(u)$ that typically do not cross grid
points. They extend along the entire grid, from the initial ray $u=u_{0}$
up to $u=u_{max}$, unless they encounter $r=0$ (or $v=v_{max}$)
before $u=u_{max}$. We derive these curves by solving the geodesic
equation and use second order interpolation in order to find the unknown
functions on these curves. In addition, we calculate the proper time
$\tau$ for each geodesic; we set $\tau=0$ on each geodesic to be
the time in which it crosses the EH.

MO and EO considered a family of radial geodesics related to each
other by time translation, or a time-translated set of geodesics (TTSG).
\footnote{In the case of a self-gravitating scalar-field perturbation described
by EO, this symmetry was merely approximated, not exact. The approximation
improved with the increase in $v$ (due to the decay in the scalar
field), up to the point in which the results could not be distinguished
from those expected on an exact set of TTSG.%
} TTSG are especially useful for gravitational shock analysis since
they share the same $r(\tau)$ function up to the location of the
shock (if exists). Hence observation of shock formation and development
is very simple. Unfortunately, time translation symmetry is a property
of static geometries and thus no longer available in our case. 

We define our family of timelike geodesics in a similar fashion to
EO approximated TTSG in the self-gravitating scalar-field case, using
an analogy to the behavior of exact radial TTSG in RN spacetime with
energy parameter $E=-u_{t}=1$. For these RN geodesics, $\dot{r}$
satisfies

\[
\dot{r}=-\sqrt{E^{2}-(1-\frac{2M}{r}+\frac{Q^{2}}{r^{2}})}\,\,,
\]

\noindent where $M$ and $Q$ are the mass and charge parameters of
RN BH. This relation reduces to $-(2M/r-Q^{2}/r^{2})^{1/2}$ in the
case of $E=1$; we replace $M$ with $m(u_{0}v)$, the initial mass
parameter at the first point of the geodesic, to obtain

\begin{equation}
\dot{r}(u_{0},v)=-\sqrt{\frac{2m(u_{0},v)}{r(u_{0},v)}-\frac{Q^{2}}{r(u_{0},v)^{2}}}\,\,\,.\label{eq:geodesics}
\end{equation}

\noindent Eq. (\ref{eq:geodesics}) defines our radial timelike geodesic
family and supplies the required initial condition for the solution
of the geodesic equation. In the self gravitating scalar field case
analyzed by EO, this definition reproduces, asymptotically, the behaviour
of an exact radial TTSG in RN spacetime. In our case it does not,
since our simulation is not asymptotically static. Our geodesic family
should not be considered an approximated TTSG by any means, but rather
as a family of radial timelike geodesics (weakly) correlated by their
initial $\dot{r}$ value.

\subsubsection{Null geodesics\label{sub:N-geo-def}}

Null geodesics are ``simpler'' than timelike geodesics since in
our double-null grid they are just $v=const$ grid rays. We do not
need to solve the geodesic equation in order to find their shape;
we do not need to interpolate our grid results; we just need to define
and calculate an appropriate affine parameter $\lambda$ and associate
it to a certain $v=const$ ray in our grid. This ``straight line''
behavior is also convenient to shock analysis since it allows a clear
observation of the gravitational shock in three-dimensional graphs. 

The derivation of the affine parameter is described in appendix A2
of EO. The affine parameter $\lambda$ has normalization freedom;
we fix this freedom by demanding $\frac{d\lambda}{dr}=-1$ at the
EH. We also set $\lambda=0$ to be the EH crossing time on each null
geodesic.

\subsubsection{Timing parameters\label{sub:T-parameter-def}}

Shock analysis requires the assignation of a timing parameter value
for each geodesic, which is particularly important for the analysis
of shock sharpening rate. In principle, a timing parameter assignment
involves two distinct choices; the choice of an appropriate time coordinate
and the choice of a specific point on the geodesic in which the coordinate
is evaluated. MO and EO used the same timing parameter for their timelike
geodesics, $v_{eh}$, the value of Eddington advanced time coordinate
at the EH crossing. %
\footnote{In the case of a self-gravitating scalar-field perturbation described
by EO, this coordinate was redefined as an Eddington-like advanced
time coordinate rather than exact Eddington coordinate, i.e. it reproduced
the expected behaviour of Eddington advanced time coordinate in RN
spacetime asymptotically.%
} Since in our scenarios we do not have a relevant RN (or asymptotically
RN) geometry to associate this coordinate with, this choice is no
longer an option. Instead, we use timing parameters based on RNV advanced
time coordinate. We use two variants of this coordinate, the initial
ray RNV advanced time coordinate, denoted $V$, and the event horizon
RNV advanced time coordinate, denoted $\widetilde{V}$. Both variants
are derived through association with the ingoing RNV metric (Eq. (\ref{eq:RNV-metric})).
The derivation of each variant and their interpretation in our different
scenarios are described at Sec. \ref{sub:v-Vaidya-def} of the appendix.

The association of such timing parameter to null geodesics is pretty
straightforward; since they are simply $v=const$ lines, they also
have a single value of $V$ or $\widetilde{V}$. The association of
timing parameter to timelike geodesics is more complex. We consider
in this paper two timing parameters for timelike geodesics, $V_{h}$,
which is the value of $V$ at the EH crossing, and $\widetilde{V}_{f}$,
which is the last $\widetilde{V}$ value of the geodesic.

\subsection{Presentation of numerical results\label{sub:Presentation}}

As we already mentioned above, we have employed several grid refinement
levels in our simulation. Most of the numerical results displayed
in this paper are based on data from the best resolution, $N=640$.
There are two exceptions: (i) The contour graphs of $r(u,v)$ (Figs.
\ref{fig:RNV-ruv}, \ref{fig:DNF-ruv}, and \ref{fig:Mixed-ruv})
are based on data from $N=640$, but there is a sampling procedure
involved in order to avoid memory problems. We specify the sampling
rate in $u$ and $v$ ($\Delta_{s}u,\Delta_{s}v$) on the caption
of each figure; we have chosen it with care in order to avoid misrepresentation
of the results. (ii) The graphs which display data along individual
timelike geodesics (Figs. \ref{fig:RNV-NoShock}(a), \ref{fig:DNF-shock-2d}(a)
and \ref{fig:Mixed-shock-2d}(a)) also display data from the second
best resolution $(N=320)$ in dashed curves (where the $N=640$ data
is displayed in solid curves). However, the results from the different
resolutions overlap, and the $N=320$ data is indistinguishable. The
shifted versions of these graphs (Figs. \ref{fig:DNF-shock-2d}(c)
and \ref{fig:Mixed-shock-2d}(c)) include just the data from $N=640$.

We use standard general relativistic units in which $c=G=1$, and
an additional unit choice which fixes the initial RN mass parameter
as $M_{0}=1$. We also set $u_{0}=v_{0}=0$ in the numerics.

\section{SINGLE NULL FLUID CASE\label{sec:RNV}}

We begin our analysis with the simplest case, a preexisting RN BH
accreting a single (ingoing) null-fluid. Although we do not observe
a gravitational shock in this case, it allows us to establish the
descriptions of the ingoing null-fluid stream and the basic shock
wave analysis before the study of more complex cases. We first describe
the setup of initial data for the initial RN geometry and $T_{vv}^{NF}$;
we move on to describe the resultant structure of spacetime and the
location of the domain of integration in it; we then demonstrate the
lack of evidence for a gravitational shock presence.

\subsection{Basic parameters and initial conditions \label{sub:IC-RNV}}

The initial RN geometry could be described in Schwarzschild coordinates
as 

\begin{equation}
ds^{2}=-f(r)dt^{2}+f(r)^{-1}dr^{2}+r^{2}d\Omega^{2}\,,\label{eq:RN-metric}
\end{equation}
where $f\equiv1-2M_{0}/r+Q^{2}/r^{2}$. Following EO choice, we choose
here an initial mass of $M_{0}=1$ (which is actually a unit choice)
and a charge parameter $Q=0.95$. 

The ingoing null-fluid stream is defined on the initial ray $u=u_{0}$;
it begins at a certain $r$ value on this ray (denoted $r_{1}$) and
does not cease up to the maximal grid ray $v=v_{max}$. The stream
has the general form

\begin{center}
\begin{equation}
T_{vv}^{NF}(u_{0},v)=\begin{cases}
\begin{array}{c}
A_{1}(1-e^{r_{1}-r})^{2}\frac{1}{r^{2}r,_{u}^{2}}\\
0
\end{array} & \begin{array}{c}
|\,\,\,\,\,\,\,\,\,\,\,\,\, r\geq r_{1}\\
|\,\, otherwise,
\end{array}\end{cases}\label{eq:Tvv_shape}
\end{equation}

\par\end{center}

\noindent where $A_{1}$ is an amplitude parameter and $r,r,_{u}$
are functions of $v$ along the initial ray. We have selected this
form to emulate the behaviour of a linear null fluid at late times;
\footnote{A linear (ingoing) null-fluid is a null fluid with linear contribution
to the mass function in RNV advanced time coordinate, $m(V)$. The
linear form is favored due to it is simplicity; it was also favored
in different models of Vaidya \cite{1986-Waugh-Lake-Linear-Mass-Vaidya}
and charged Vaidya \cite{Bonnor-Vaidya-Sings1,Bonnor-Vaidya-Sings2}
spacetimes due to the self-similar nature it allows in these models
and the ability to derive the inner structure analytically.%
} we have achieved this goal, as demonstrated in Sec. \ref{sub:Linear-mass-calc}
of the appendix. The factor $(1-e^{r_{1}-r})^{2}$ ensures a smooth
(and quick) transition from the null phase to the linear phase. We
choose $r_{1}=10$; this value is reached on the initial ray $u=u_{0}$
at $v_{3}\approx12.54$ (or $V_{3}\approx5.63$). Although the stream
does not cease in our simulation, we assume it ceases at some point
in the future, denoted $v=v_{4}$ (or $V=V_{4})$, so that the asymptotic
spacetime has RN geometry with a new (and unknown) mass parameter
$m_{final}$ and a charge parameter $Q=0.95$. %
\footnote{Note that this choice introduces some approximation on our derivation
of the EH location (see Sec. \ref{sub:Mass-function}); the actual
$u$ value of the EH is expected to be lower than $u_{h}$. However,
as $v_{4}$ is undetermined, one could take it to be arbitrarily close
to $v_{max}$, so the deviation could be small.%
} The amplitude $A_{1}$ monitors the mass contribution of the stream;
we choose $A_{1}=3.4555203\times10^{-4}$ which yields a mass of $m_{BH,vmax}=2.5$.
This mass fits EH value of $r_{+,vmax}\simeq4.812$, IH value of $r_{-,vmax}\simeq0.1875$,
and an extreme value of the IH surface gravity at $v=v_{max},$ $\kappa_{-,vmax}$$\simeq65.8$. 

The remaining initial values are taken according to the first column
of table \ref{tab:IC-def}. In particular, $T_{vv}^{NF}$ vanishes
on $v=v_{0}$; the outgoing null fluid $T_{uu}^{NF}$ and the scalar
field $\Phi$ vanish on both initial rays $(u=u_{0}$ and $v=v_{0}$);
$\sigma$ conforms on both rays with the maximal-$\sigma$ gauge condition
(Eq. (\ref{eq: Sigma_max:u+v})). $r$ is calculated numerically on
both rays from the solution of the constraint equations, Eq. (\ref{eq:  ruu})
at $v=v_{0}$ and Eq. (\ref{eq:  rvv}) at $u=u_{0}$.

The domain of integration is $u_{0}=v_{0}=0$, $u_{max}=260.5421875$,
\footnote{The values of $u_{max}$ are typically not round due to the $r(u,v_{0})$
cutoff near the singularity (at $r=0.1$). See Sec. \ref{sec:Numerical-Algorithm}.%
} and $v_{max}=120$. The value of $r$ in the initial vertex is $r_{0}=5$
and it grows up to $r\simeq49.47$ at $r(u_{0},v_{max})$. The other
two corners of the grid are $r(u_{max},v_{0})\simeq0.1004$ and $r(u_{max},v_{max})\simeq0.1868$.

\subsection{The structure of spacetime\label{sub:ST-RNV}}

The location of the numerical grid in spacetime and the structure
of spacetime are illustrated in Fig. \ref{fig:Penrose-RNV}. The outgoing
initial ray $u=u_{0}$ is located outside the BH, while the ingoing
initial ray $v=v_{0}$ penetrates the BH and passes the outgoing IH.
Panel (b) demonstrates that spacetime could be divided into three
distinct patches: (i) the initial RN geometry (RN1, at $v<v_{3}$)
with mass parameter $M_{0}$ and charge $Q$; (ii) the ingoing null-fluid
stream region ($v_{3}\leq v\leq v_{4})$, denoted RNVi; (iii) the
final RN Geometry (RN2, at $v>v_{4})$, with mass parameter $m_{final}$
and charge $Q$, which is not covered by our simulation. All three
patches of spacetime are extendable; the only ``neighboring'' singularity
is the original timelike $r=0$ singularity of the initial RN geometry.
\footnote{There are several consistent choices and assumptions involved in the
drawing of spacetime diagrams in this paper (including Fig. \ref{fig:Penrose-SGSF}).
We assume that the initial RN geometry belongs to an eternal RN spacetime
for the sake of simplicity; this geometry could be a result of a (more
physical) collapse scenario as well. The past ingoing EH in the diagrams
belongs to this initial RN geometry, where the outgoing EH and IH
and the ingoing IH belong to the final RN (or perturbed charged) geometry.
We also assume there are no additional perturbations in spacetime
besides those defined on our domain of integration, although we do
extend these perturbations backward in time to their source (either
null infinity or the ingoing EH).%
} (There are also similar singularities of RNV geometry and the final
RN geometry, but they are not drawn in this diagram since they are
not in the vicinity of the grid).

This structure is demonstrated by the numerical results of $r(u,v)$,
depicted as a contour graph in Fig. \ref{fig:RNV-ruv}. Panel (a)
describes the entire grid; in particular, one could notice the zone
outside the BH as a region in which $r$ values rise with $v$ (at
the bottom of the graph, up to $u_{h}\simeq69.29$). Panel (b) focuses
at the vicinity of $u=u_{max}=260.5421875$. This high zoom level
(of order of $0.2$ in $u$) demonstrates that the grid ends regularly. 

\begin{center}
\begin{figure}[H]
\noindent \begin{centering}
\includegraphics[scale=0.5]{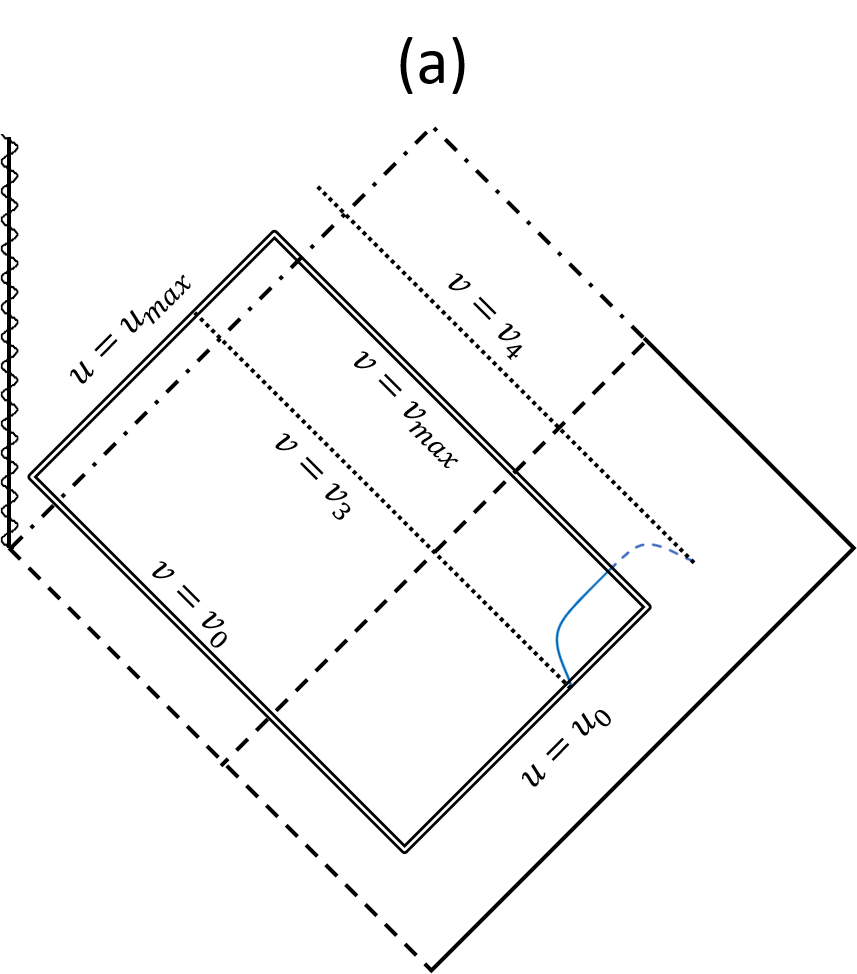}\hspace{2cm}\includegraphics[scale=0.5]{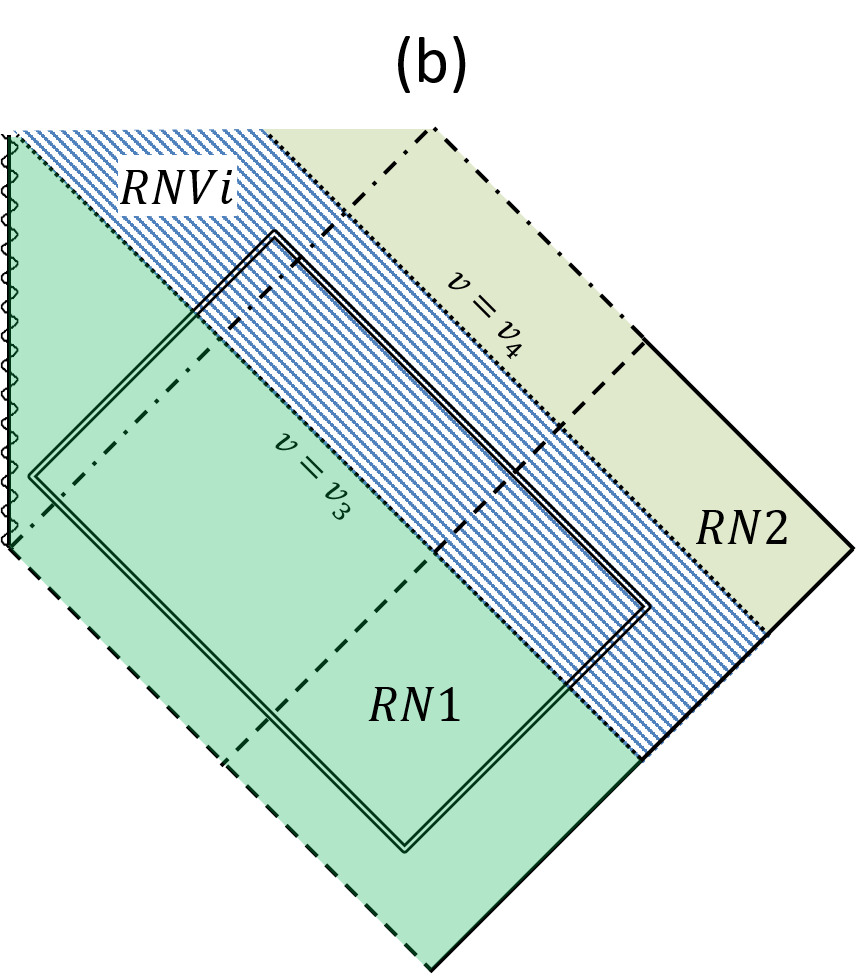}
\par\end{centering}

\protect\caption{\label{fig:Penrose-RNV} Penrose diagrams illustrating the structure
of spacetime in the single null fluid case. Both panels describe the
location of the numerical domain of integration and the ingoing null
fluid stream in spacetime; while panel (a) focuses on initial data,
panel (b) analyzes the different patches of spacetime. In both panels,
the domain's limits are denoted by a double black line; solid black
lines denote null infinity; dashed black lines denote the EH; dashed-dotted
black lines denote the IH. Wavy vertical lines denote the timelike
$r=0$ singularity of RN geometry; dotted black lines denote the ingoing
null-fluid stream limits $v=v_{3}$ and $v=v_{4}$. The blue curve
in panel (a) represents the ingoing null-fluid stream on the initial
ray $u=u_{0}$; its shape roughly describes the stream contribution
to the mass function derivative $m,_{V}$. The dashed part of the
curve represents the extension of the stream outside the domain of
integration up to its termination at $v=v_{4}$. Panel (b) demonstrates
that spacetime could be divided into three distinct patches; patches
RN1 and RN2 fit the initial RN geometry and the final RN geometry
accordingly; they are separated by the ingoing null-fluid stream patch
RNVi (diagonal blue stripes). }
\end{figure}

\par\end{center}

\begin{center}
\begin{figure}[H]
\noindent \begin{centering}
\includegraphics[scale=0.35]{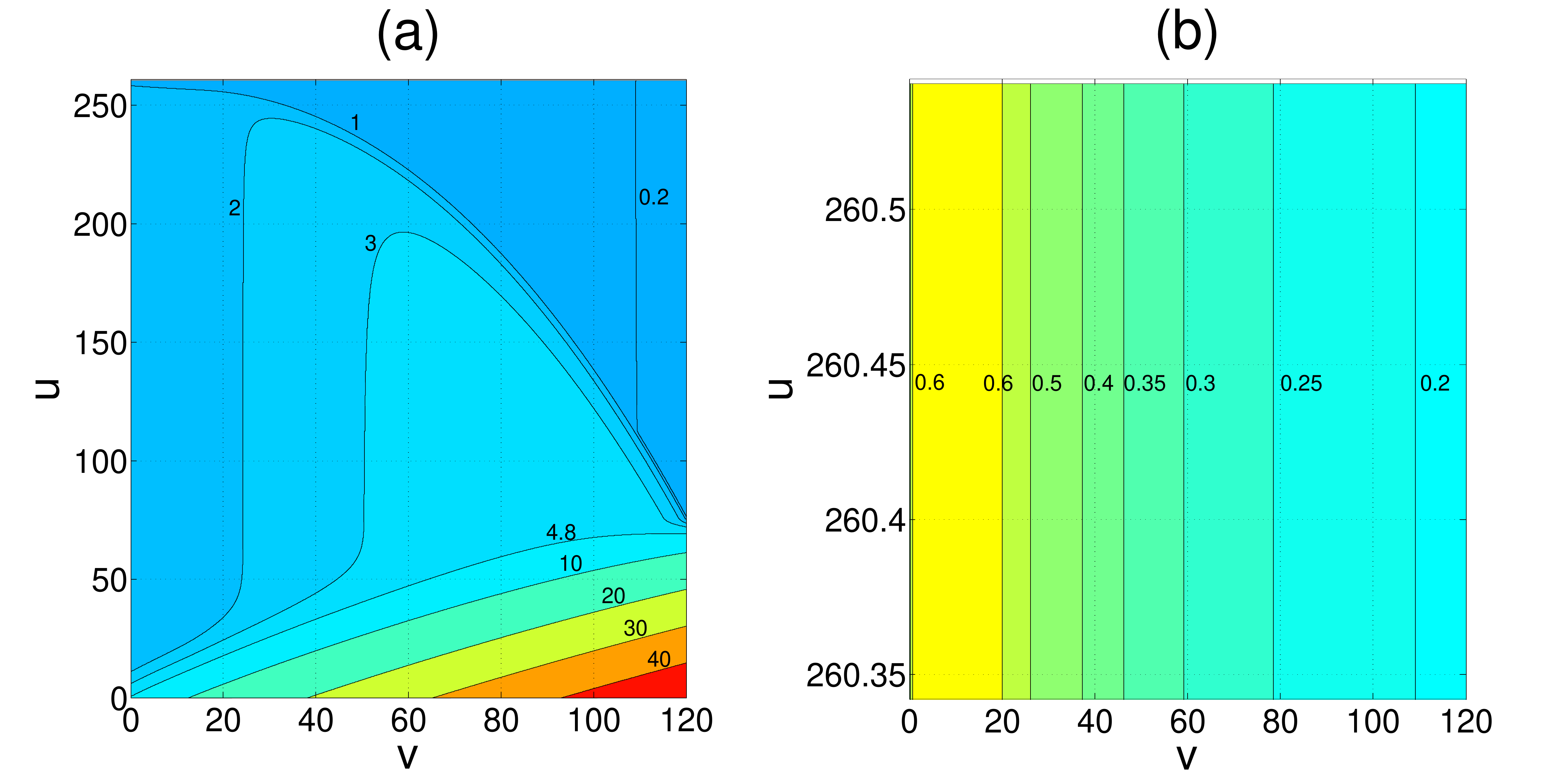}
\par\end{centering}

\protect\caption{\label{fig:RNV-ruv} Numerical results for $r(u,v)$ in the single
null fluid case. Both panels display contour graphs of $r(u,v)$ based
on numerical results; panel (a) displays results on the entire grid
and panel (b) displays a zoom deep inside the BH, near $u=u_{max}=260.5421875$.
The panels use different color code and different level choices for
$r$ for the sake of visibility. The zoom in panel (b) reveals that
the grid ends regularly. Panel (a) is based on the results of $N=640$
sampled in a coarse resolution $(\Delta_{s}u=\Delta_{s}v=0.1)$, while
panel (b) is based on $N=640$ results with a sampling on $v$ alone
$(\Delta_{s}v=0.05)$. }
\end{figure}

\par\end{center}

\subsection{Gravitational shock absence \label{sub:Shock-RNV}}

The lack of evidence for a gravitational shock is demonstrated in
Fig. \ref{fig:RNV-NoShock}. Panel (a) displays numerical results
for $r(\tau)$ on a family of ingoing radial timelike geodesics; panel
(b) displays similar results for $r(\lambda)$ on a series of ingoing
radial null geodesics, or grid rays. In both cases, $r(\tau)$/$r(\lambda)$
is a smooth curve for all the geodesics; we see it drops to the expected
IH value (the dashed black curve) and ``freezes'' there (the earliest
geodesics, e.g. the timelike $V_{h}=0$ or the null $V=-3.2$ actually
pass this value). 

Due to the lack of time translation symmetry, the geodesics reaches
their corresponding (different) $r_{-}$ values at different values
of $\tau/\lambda$. This characteristic impedes shock observation
due to ``scale stretching''; we deal with this problem in the next
section. 

\begin{center}
\begin{figure}[H]
\noindent \begin{centering}
\includegraphics[scale=0.35]{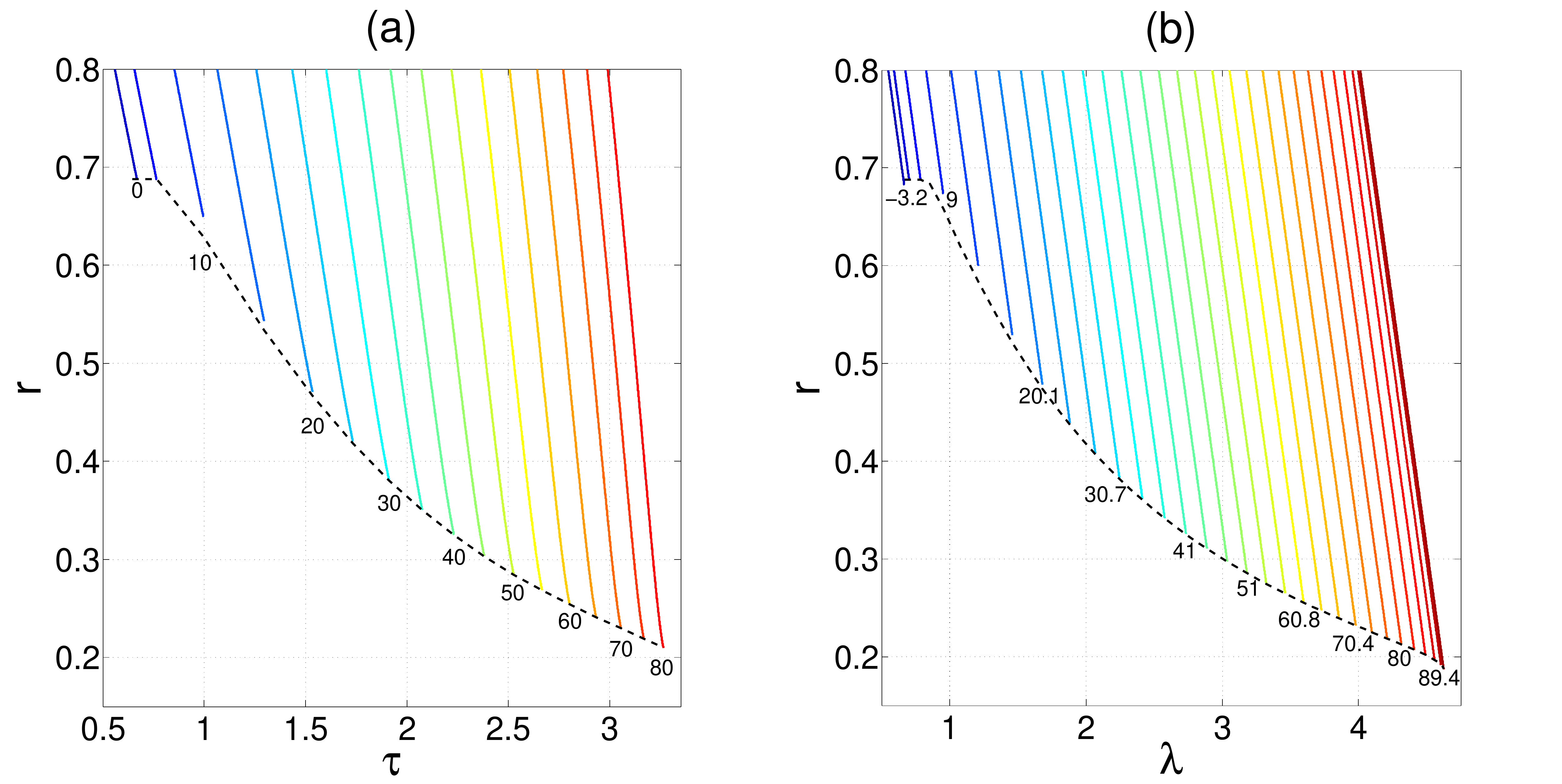}
\par\end{centering}

\protect\caption{\label{fig:RNV-NoShock} Gravitational shock absence in the single
null fluid case. Panel (a) presents $r$ as a function of proper time
$\tau$ along a family of ingoing radial timelike geodesics; panel
(b) presents $r$ as a function of affine parameter $\lambda$ along
a series of ingoing radial null geodesics. In both panels, geodesics
are solid lines, distinguished by different colors and different timing
parameter values; the timing parameter is $V_{h}$ (the value of initial
ray RNV advanced time at the EH) for timelike geodesics and $V$ for
null geodesics. Panel (a) contains timelike geodesics at the range
$0\leq V_{h}\leq80$, where $V_{h}$ increases by increments of $5$
from left $(V_{h}=0$) to right $(V_{h}=80)$; panel (b) contains
grid null rays at the range $-3.3<V<95.7$, with irregular $V$ values
and irregular increments of $V$ from left $(V\simeq-3.2)$ to right
$(V\simeq95.6)$. The dashed black curve denotes the corresponding
IH value of the geodesics; $r_{-}(V_{h})$ in panel (a) and $r_{-}(V)$
in panel (b). Gravitational shock absence manifests in the smooth
shape of $r(\tau)/r(\lambda)$.}
\end{figure}

\par\end{center}

\section{TWO NULL FLUIDS CASE \label{sec:DNF}}

We construct our two null fluids case through the inclusion of an
outgoing null fluid pulse in the initial data setup described in Sec.
\ref{sub:IC-RNV}. In addition to demonstrating the existence of a
gravitational shock wave, we uncover some interesting results on the
inner structure: a strong evidence for the existence of a spacelike
$r=0$ singularity, and a possible indication for the existence of
a null, non naked, $r=0$ singularity. The structure of this section
is similar to the previous section with one exception: since our shock
analysis has positive results, we provide an additional analysis of
the shock sharpening rate, and compare it to MO and EO's result.

\subsection{Basic parameters and initial conditions \label{sub:IC-DNF}}

The initial spacetime in this case is the same as in the previous
section --- RN spacetime with mass parameter $M_{0}=1$ and charge
parameter $Q=0.95$. The ingoing null fluid stream is also identical
to the stream described in section \ref{sub:IC-RNV}; it has the same
initial function $T_{vv}^{NF}$ on the initial outgoing ray $u=u_{0}$
(Eq. (\ref{eq:Tvv_shape})) and the same parameters (begins at $r_{1}=10/v_{3}\approx12.54/V_{3}\approx5.63$
and has an amplitude of $A_{1}=3.4555203\times10^{-4}$). However,
we now include an outgoing null-fluid pulse, defined on the initial
ingoing ray $v=v_{0}$ by

\begin{center}
\begin{equation}
T_{uu}^{NF}(u,v_{0})=\begin{cases}
\begin{array}{c}
A_{2}\frac{16(r_{a}-r)^{3}(r-r_{b})^{3}}{\pi(r_{a}-r_{b})^{6}}r,_{u}^{2}\\
0
\end{array} & \begin{array}{c}
|\,\,\, r_{b}\leq r\leq r_{a}\\
|\,\,\,\, otherwise.
\end{array}\end{cases}\label{eq:Tuu_Shape}
\end{equation}

\par\end{center}

\noindent where $A_{2}$ is an amplitude parameter and $r,r,_{u}$
are functions of $u$ along the initial ray. The polynomial form has
compact support and is limited to a certain $r$ range ($r_{b}\leq r\leq r_{a}$,
which translates to a certain $u$ range, $u_{1}\leq u\leq u_{2}$)
in order to control the mass function. The factor $r,_{u}^{2}$ allows
us to avoid a numerical problem of ``mass blow up'' at the EH, a
possible hazard of our gauge choice. %
\footnote{\noindent Due to our gauge selection (Eq. (\ref{eq: Sigma_max:u+v})),
$r$ value is frozen on the initial ray $v=v_{0}$ at the vicinity
of the EH and $r,_{u}$ vanishes there (as explained in Ref. \cite{Eilon-Ori_AdaptiveGauge}).
If $T_{uu}^{NF}(u_{h},v_{0})$ is nonvanishing we would have diverging
pulse mass at the EH.%
} We choose the amplitude as $A_{2}=12.0$, and the pulse limits as
$r_{a}=1.54$ and $r_{b}=0.7$, which translates to $u_{1}\simeq16.47$
and $u_{2}\simeq253.72$. Since the EH is located at $u_{h}\simeq69.29$
\footnote{\noindent The difference between this value and the one in the single
null fluid case is of order $10^{-3}$.%
}, the outgoing pulse begins well outside the BH and ends deep inside
the BH; the location of the pulse and its general shape are illustrated
in panel (a) of Fig. \ref{fig:Penrose-DNF}. Although the pulse has
a simple shape in $r$, the factor $r,_{u}^{2}$ dampens it greatly
at the vicinity of the EH; as a result, the effect of the pulse outside
the BH turns out to be rather minor, while the effect inside the BH
is much more pronounced, as demonstrated below. However, the outgoing
pulse does decrease the mass of the BH, as now $m_{BH,vmax}=2.4925$.
This mass fits EH value of $r_{+,vmax}\simeq4.797$, IH value of $r_{-,vmax}\simeq0.1881$,
and surface gravity value of $\kappa_{-,vmax}$$\simeq65.1$. 

The remaining initial values are taken according to the second column
of table \ref{tab:IC-def}. In particular, $T_{vv}^{NF}$ vanishes
on $v=v_{0}$; $T_{uu}^{NF}$ vanishes on $u=u_{0}$; the scalar field
$\Phi$ vanishes on both initial rays and $\sigma$ conforms on both
rays with the maximal-$\sigma$ gauge condition (Eq. (\ref{eq: Sigma_max:u+v})).

The domain of integration is $u_{0}=v_{0}=0$, $u_{max}=253.7375$,
and $v_{max}=120$. The value of $r$ in the initial vertex is $r_{0}=5$
and it grows up to $r\simeq49.47$ at $r(u_{0},v_{max})$. The third
corner of the grid is $r(u_{max},v_{0})\simeq0.1326$; however, there
are no numerical results available in the fourth corner of the grid
$(u_{max},v_{max})$ due to the presence of a spacelike singularity.

\subsection{The structure of spacetime\label{sub:ST-DNF}}

Fig. \ref{fig:Penrose-DNF}(b) reveals the complex structure of spacetime
--- it consists of eight patches of different effective spacetime
(although only six are covered in our simulation). The initial RN
patch (RN1) borders an outgoing null-fluid patch (RNVo) at $u=u_{1}$
and an ingoing null-fluid patch (RNVi) at $v=v_{3}.$ There are two
additional RN patches: the asymptotic RN patch (RN2, at $u<u_{1}$
and $v>v_{4}$) is not covered in our simulation and has unknown mass
parameter $m_{final}$; the inner RN patch (RN3, at $u>u_{2}$ and
$v<v_{3}$) is located deep inside the BH (after the outgoing null-fluid
pulse) and has mass parameter $m_{3}\approx1.569$. %
\footnote{Due to the proximity of $u_{2}\simeq253.72$ to $u_{max}=253.7375$,
this value is somewhat uncertain. The numerical error is of order
$0.002$ but convergence quality is poor.%
} The patch where the fluids intersect (TNF, at $u_{1}\leq u\leq u_{2}$,
$v_{3}\leq v\leq v_{4})$ contains the shock wave and a spacelike
$r=0$ singularity; we also see a possible indication for a section
of null $r=0$ singularity, as we discuss below. There are additional
single fluid patches where the null fluids separate, which are (in
principle) extendable. 

At first glance, the numerical results for $r(u,v)$ (Fig. \ref{fig:DNF-ruv})
appears quite similar to the single null fluid case; panel (a), which
displays results on the entire grid, seems almost identical to panel
(a) of Fig. \ref{fig:RNV-ruv}. A zoom near $u=u_{max}=253.7375$
(panels (b) and (c)) reveals a fundamental difference; while in the
single null fluid case the grid ended regularly, in case of two null
fluids we detect a $r=0$ singularity which breaks down the numerics
(the border with the criss-cross patch, in which results are unavailable).
The exact nature of this singularity is somewhat open to debate. While
it clearly contains a spacelike section --- the diagonal border of
the criss-cross patch in panel (b), which extends from $v\sim22$
up to $v_{max}=120$ --- it may also contain a null ($v=const$) section,
which is indicated at the top left of panel (b) and the focus of panel
(c). If this is indeed a null section and not a different spacelike
section, it is not well known; this is not a naked singularity which
is a known phenomenon in Vaidya \cite{1986-Waugh-Lake-Linear-Mass-Vaidya}
and charged Vaidya \cite{Bonnor-Vaidya-Sings1,Bonnor-Vaidya-Sings2}
geometries. The null classification is uncertain, however, mainly
due to the proximity to the edge of the grid (where numerical fluxes
may cause unexpected effects) %
\footnote{For instance, we suspect that the slight turn left of this section
of the singularity at $u=u_{max}=253.7375$ is an artifact due to
numerical fluxes.%
} and the short span of this section ($\sim0.03$ in $u$). We also
note that despite the way this section is drawn in Fig. \ref{fig:Penrose-DNF},
it is uncertain if this section begins at the end of the outgoing
null fluid pulse $(u_{2}\simeq253.72)$ or earlier. We conclude that
this section (and the nature of singularities in this case in general)
requires further study, which is beyond the scope of the current paper. 

\begin{center}
\begin{figure}[H]
\noindent \begin{centering}
\includegraphics[scale=0.5]{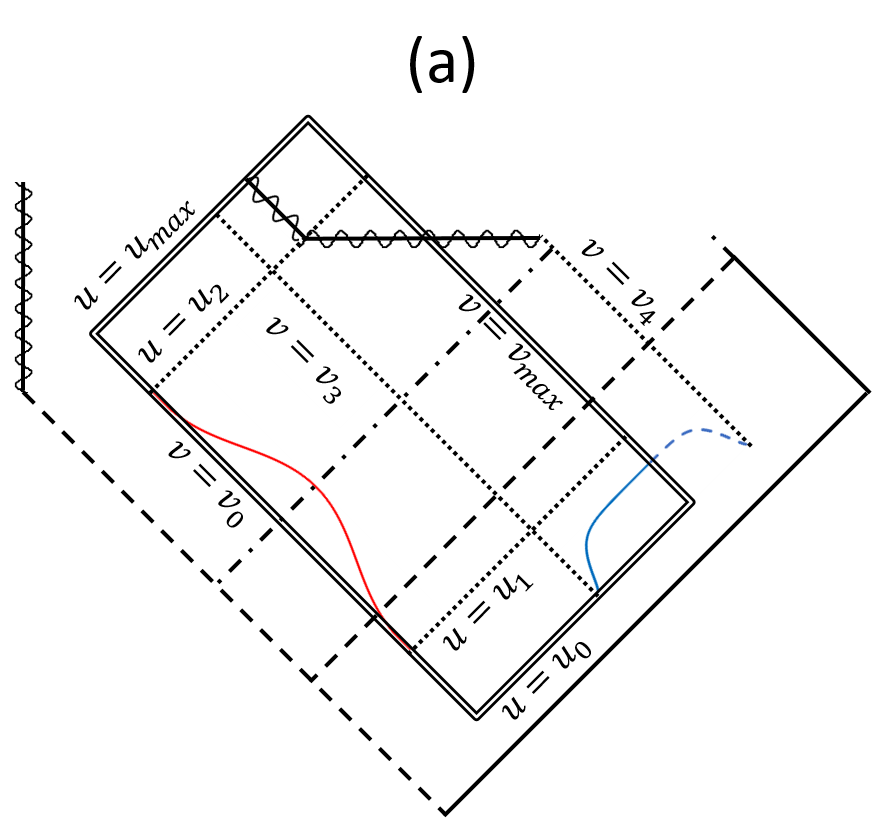}\hspace{2cm}\includegraphics[scale=0.5]{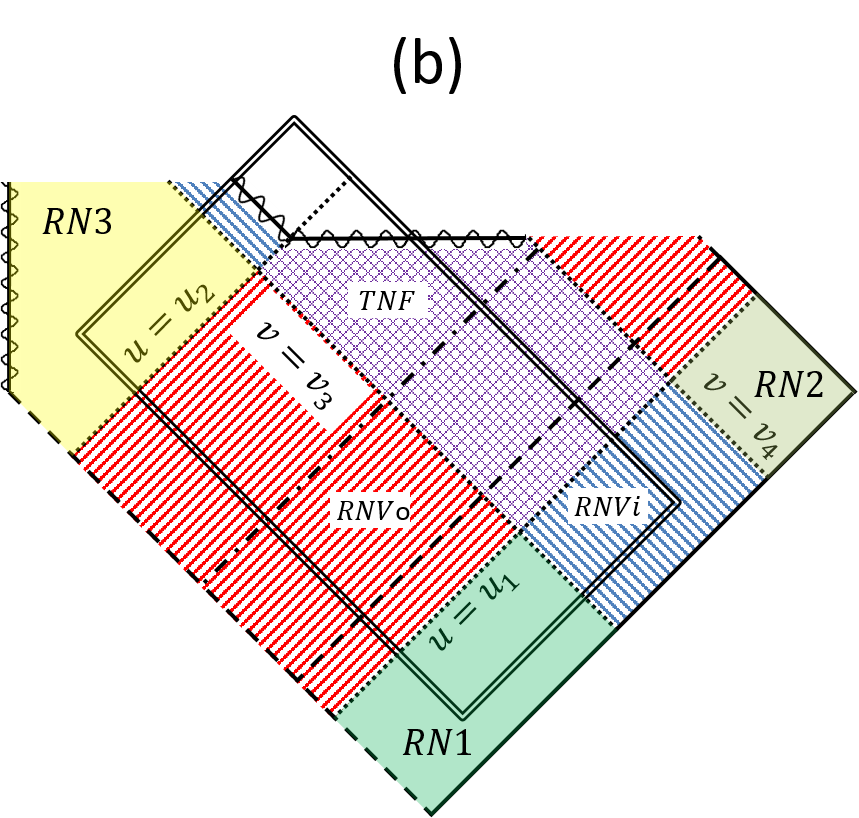}
\par\end{centering}

\protect\caption{\label{fig:Penrose-DNF} Penrose diagrams illustrating the structure
of spacetime in the case of two null fluids. Both panels describe
the location of the numerical domain of integration and the null fluids
in spacetime; while panel (a) focuses on initial data, panel (b) analyzes
the different patches of spacetime. In both panels, the domain's limits
are denoted by a double black line; solid black lines denote null
infinity; dashed black lines denote the EH; dashed-dotted black lines
denote the IH. Wavy lines denote the timelike $r=0$ singularity of
RN geometry, as well as the spacelike $r=0$ singularity and the (suspected)
null $r=0$ singularity that develop in this case. Dotted black lines
denote the ingoing null-fluid stream limits $v=v_{3}$ and $v=v_{4}$,
as well as the outgoing null-fluid pulse limits $u=u_{1}$ and $u=u_{2}$
. The blue curve in panel (a) represents the ingoing null-fluid stream
on the initial ray $u=u_{0}$. The red curve represents the outgoing
null-fluid pulse on the initial ray $v=v_{0}$; its shape roughly
reflects the shape of the pulse in $T_{uu}$ but with a different
choice for the coordinate $u$ (say, $u'=-r(u,v_{0})$), which simplifies
its form. Panel (b) demonstrates that spacetime could be divided into
eight distinct patches: three patches of RN geometry (RN1, RN2 and
RN3, initial, asymptotic and inner accordingly), two patches of ingoing
null-fluid stream (RNVi, diagonal blue stripes), two patches of outgoing
null-fluid pulse (RNVo, diagonal red stripes), and the patch in which
both fluids intersect and the spacelike singularity develops (TNF,
the checkered purple patch).}
\end{figure}

\par\end{center}

\begin{center}
\begin{figure}[H]
\noindent \begin{centering}
\includegraphics[scale=0.35]{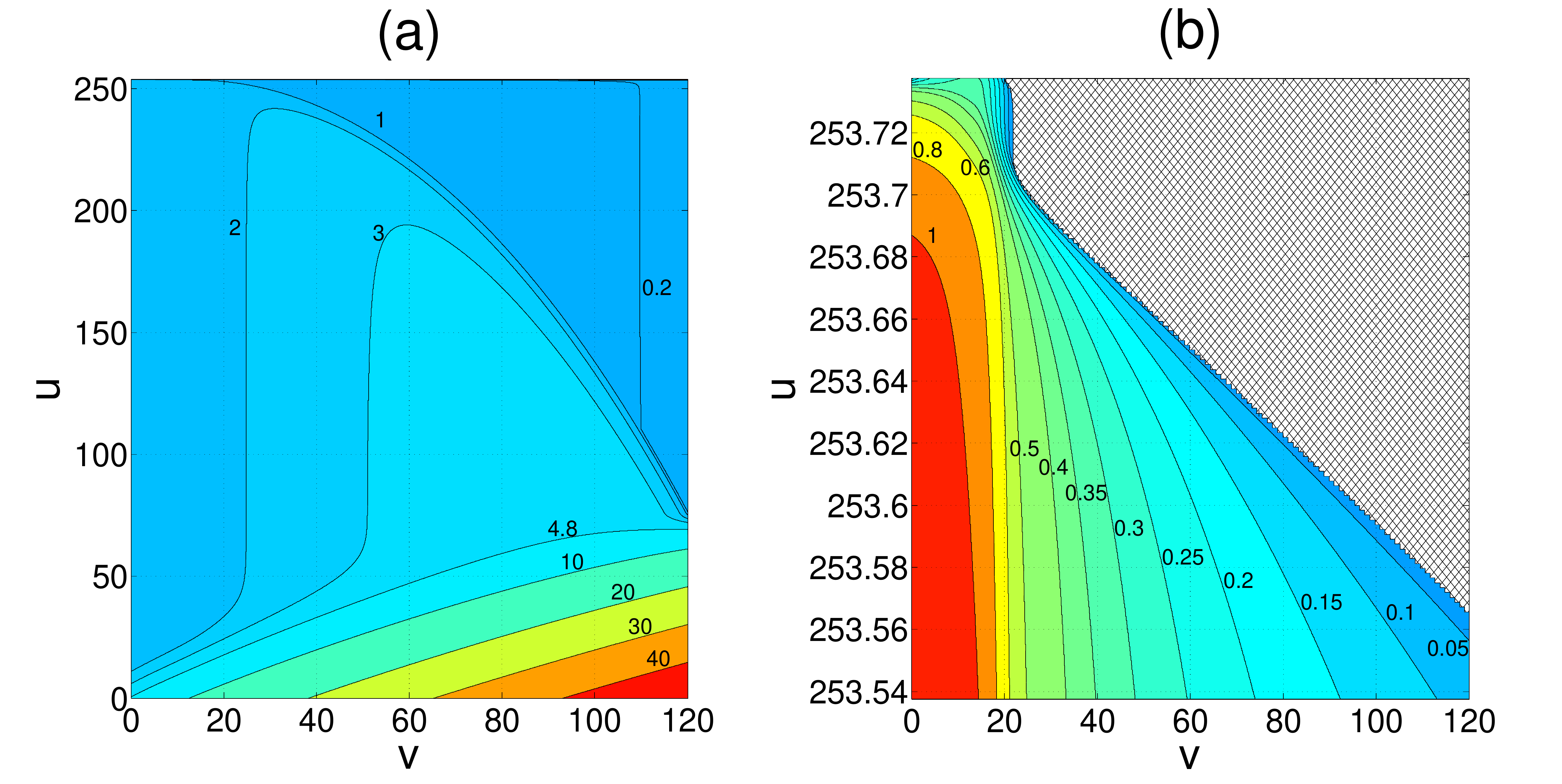}
\par\end{centering}

\noindent \begin{centering}
\includegraphics[bb=658bp 0bp 1428bp 719bp,clip,scale=0.35]{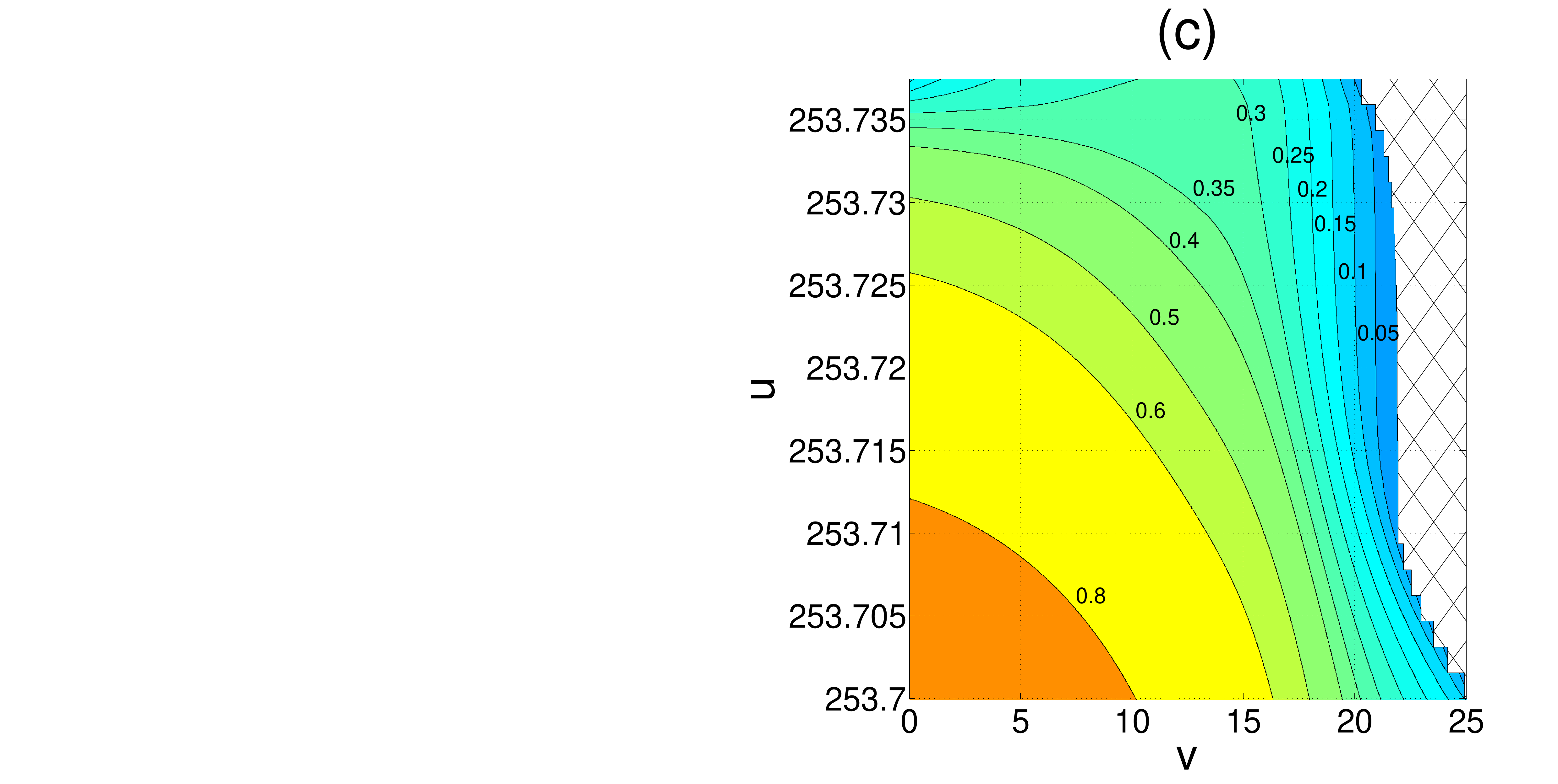}
\par\end{centering}

\protect\caption{\label{fig:DNF-ruv} Numerical results for $r(u,v)$ in the case of
two null fluids. The panels display contour graphs of $r(u,v)$ based
on numerical results; panel (a) displays results on the entire grid;
panel (b) displays a zoom deep inside the BH, near $u=u_{max}=253.7375$;
panel (c) displays a further zoom on the top left corner of panel
(b). Panel (a) uses different color code and different level choice
for $r$ from panels (b) and (c) for the sake of visibility. The criss-cross
patch on panels (b) and (c) represents a region in which numerical
results are unavailable due to the $r=0$ singularity. Panel (b) reveals
the presence of a spacelike $r=0$ singularity; it manifests as the
diagonal border of the criss cross patch. Panels (b) and (c) hints
on the presence of a null $r=0$ singularity, represented by the (approximately)
vertical border of the criss-cross patch. Panel (a) is based on $N=640$
numerical results sampled in a lower resolution $(\Delta_{s}u=\Delta_{s}v=0.1)$,
while panels (b) and (c) are based on $N=640$ numerical results without
any sampling. }
\end{figure}

\par\end{center}

\subsection{Gravitational shock detection \label{sub:Shock-DNF}}

The gravitational shock wave is demonstrated in Figs. \ref{fig:DNF-shock-2d}
and \ref{fig:DNF-shock-3d}. Panel (a) of Fig. \ref{fig:DNF-shock-2d}
presents $r(\tau)$ on a family of timelike geodesics; panel (b) displays
$r(\lambda)$ on a series of null geodesics. In both cases the shock
manifests as a sharp ``break'' or a vertical ``drop'' in $r(\tau)$/$r(\lambda)$,
located at the specific crossing point of $r=r_{-}$ of the geodesic.
Since the aforementioned ``scale stretching'' problem prevents a
close inspection of the break (and obscures the shock), we draw the
geodesics together by a simple shift in $r$ and $\tau/\lambda$ in
panels (c) and (d) of Fig. \ref{fig:DNF-shock-2d}. These panels display
the same geodesics as panels (a) and (b) (accordingly), but each curve
is now shifted by a different constant $r_{-}$ and a different constant
$\tau/\lambda$; the shift by $r_{-}$ allows us to confirm more clearly
that the shock is indeed located at the specific $r_{-}$ value of
each geodesic. 

The shock does not form immediately after the null fluid stream begins
(at $V\simeq5.63$). The first clear observation of the shock is seen
in the timelike geodesic $V_{h}=20$ (Fifth from the left in panels
(a)/(c)) or in the null geodesic $V=20.1$ (Fourth from the left in
panels (b)/(d)). %
\footnote{We emphasize that $V_{h}$ and $V$ are different timing parameters,
even though they are based on the same coordinate; a timelike geodesic
crosses the horizon at a certain $V=V_{h}$ value but reaches $r=r_{-}$
at a higher, $V>V_{h}$, value . So the timelike geodesic $V_{h}=20$
is actually ``later'' in terms of shock development than the null
geodesic $V=20$, even though the geodesics intersect at the EH. Also,
we note that it appears for both geodesics (timelike $V_{h}=20$ and
null $V=20.1$, in panels (c) and (d)) like the shock begins ``prematurely'',
at $r>r_{-}$. This is a scale artifact which vanishes with a zoom
in; the geodesics are actually still smooth at the apparent breaking
point.%
} However, panels (c) and (d) indicate that once the shock forms it
sharpens very quickly; for instance, it is hard to differentiate between
the sharp features of timelike geodesics in the range $30\leq V_{h}\leq70$
in panel (c). 

While Fig. \ref{fig:DNF-shock-2d} demonstrates the shock very clearly
(especially panels (c) and (d)), the visual effect remains somewhat
limited: isolated and localized. We wish to reconstruct EO's more
global ``vertical wall'' representation of the shock. For this purpose,
Fig. \ref{fig:DNF-shock-3d} presents $r(\lambda)$ of a series of
null geodesics in a three-dimensional graph. The geodesics are taken
from a similar $V$ range to the one in Fig. \ref{fig:DNF-shock-2d}(b)
but with a denser sampling. Panel (a) presents the exact $r(\lambda)$
curves of the geodesics; panel (b) presents the same geodesics, shifted
by constant $r_{-}$ and $\lambda$ values in a similar fashion to
panels (c) and (d) of Fig. \ref{fig:DNF-shock-2d}. Although the earliest
geodesics are smooth, we can see in panel (a) that each late geodesic
``breaks'' to a sheer drop at a different $r$ value. In panel (b)
we confirm that this value is indeed the appropriate $r_{-}(V)$ value
of the geodesic; we see the formation of the infamous ``vertical
wall''.

\begin{center}
\begin{figure}[H]
\noindent \begin{centering}
\includegraphics[scale=0.35]{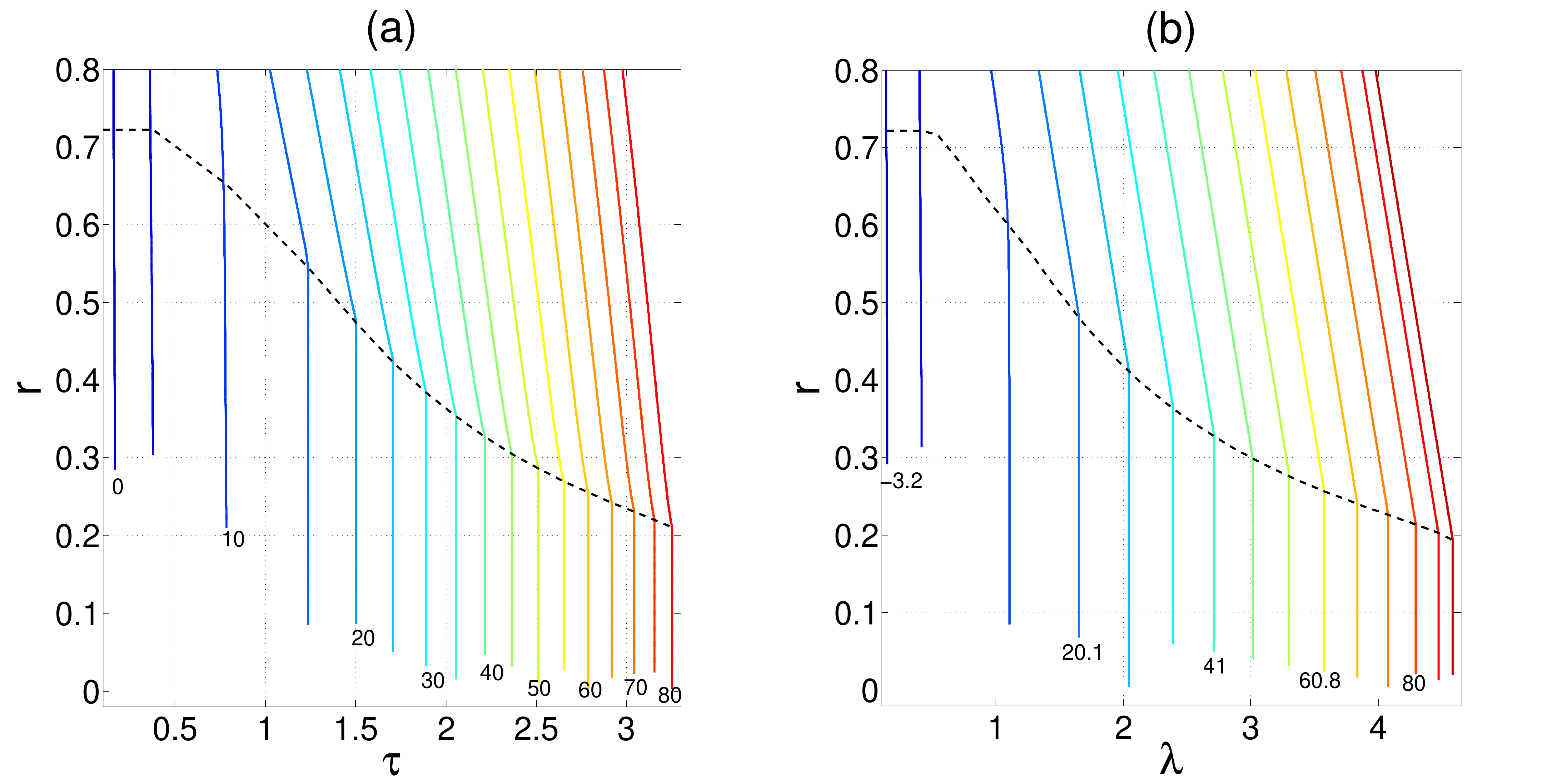}
\par\end{centering}

\noindent \begin{centering}
\includegraphics[scale=0.35]{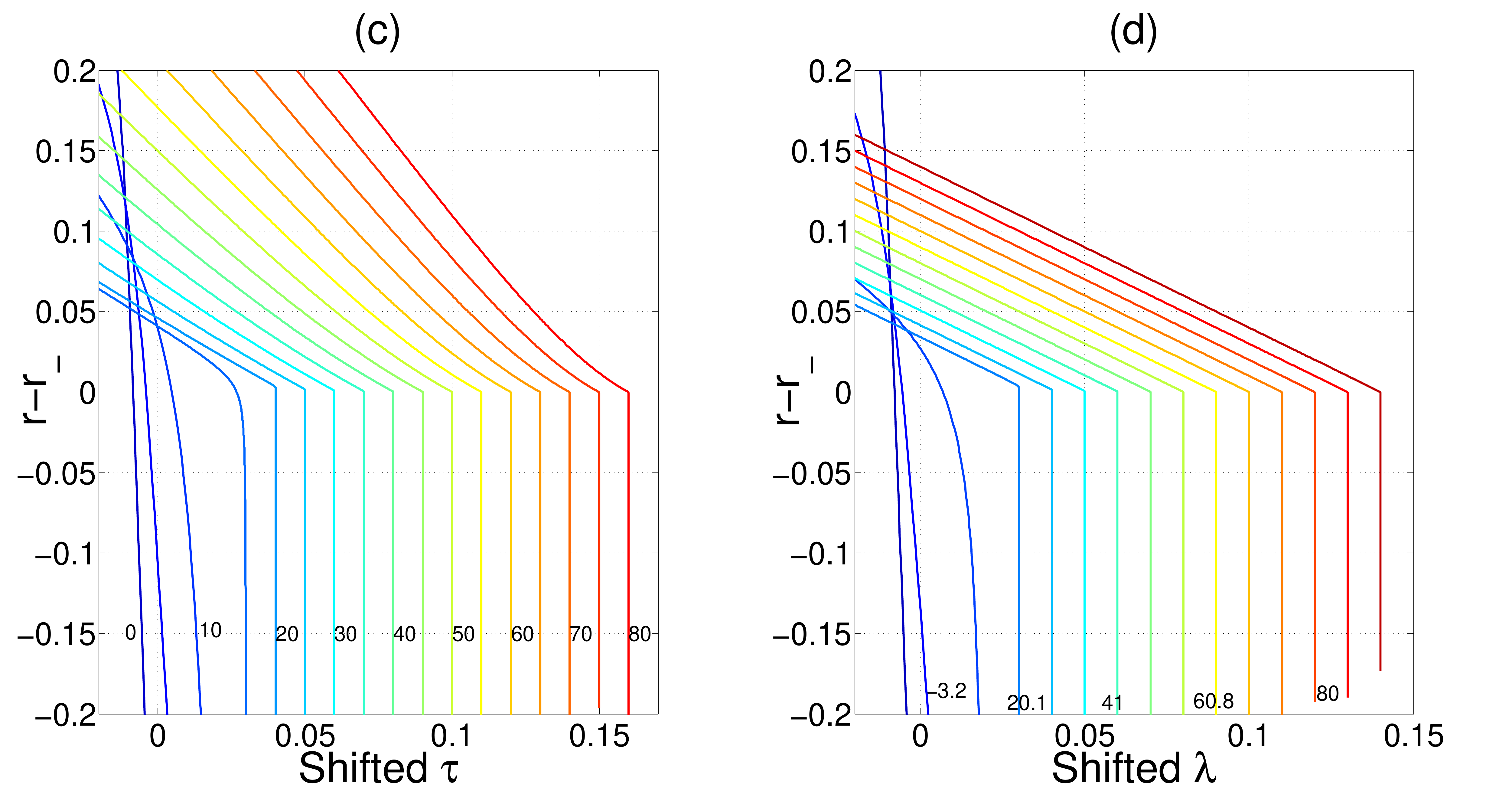}
\par\end{centering}

\protect\caption{\label{fig:DNF-shock-2d} Gravitational shock wave in case of two
null fluids. Panel (a) presents $r(\tau)$ for a family of radial
timelike geodesics; panel (b) presents $r(\lambda)$ for a series
of ingoing radial null geodesics. Geodesics are solid lines, distinguished
by different colors and different timing parameter values; the timing
parameter is $V_{h}$ for timelike geodesics and $V$ for null geodesics.
The dashed black curve in panels (a) and (b) denotes the evolving
IH value of the geodesics; $r_{-}(V_{h})$ in panel (a) and $r_{-}(V)$
in panel (b). Panel (c) displays the same timelike family of panel
(a), but shifted in $r$ and $\tau$ {[}by subtracting the specific
$r_{-}(V_{h})$ value of each geodesic from $r(\tau)$ and shifting
$\tau$ by a constant factor ($\tau_{shift}=(i-1)\times0.01-\tau_{f}^{(i)}$
for the $i$-th geodesic from the left, where $i=1,...17$ and $\tau_{f}^{(i)}$
is the last proper time value of the geodesic){]}. Panel (d) displays
the same null series of panel (b), but shifted in $r$ and $\lambda$
in a similar fashion {[}subtraction of the specific $r_{-}(V)$ value
of each geodesic from $r(\lambda)$ and a shift of $\lambda$ by $\lambda_{shift}=(i-1)\times0.01-\lambda_{f}^{(i)}$
for the $i$-th geodesic from the left, where $i=1,...15$ and $\lambda_{f}^{(i)}$
is the last affine parameter value of the geodesic). Panels (a) and
(c) contain timelike geodesics at the range $0\leq V_{h}\leq80$,
where $V_{h}$ increases by increments of $5$ from left $(V_{h}=0$)
to right $(V_{h}=80)$; panels (b) and (d) contain grid null rays
at the range $-3.3<V<92.6$, with irregular $V$ values and irregular
increments of $V$ from left $(V\simeq-3.2)$ to right $(V\simeq92.5)$.
Gravitational shock manifests as a sharp drop or ``break'' in $r(\tau)/r(\lambda)$
at the IH; the shock is first seen clearly on $V_{h}=20/V=20.1$,
and is easier to observe in panels (c) and (d).}
\end{figure}

\par\end{center}

\begin{center}
\begin{figure}[H]
\noindent \begin{centering}
\includegraphics[scale=0.35]{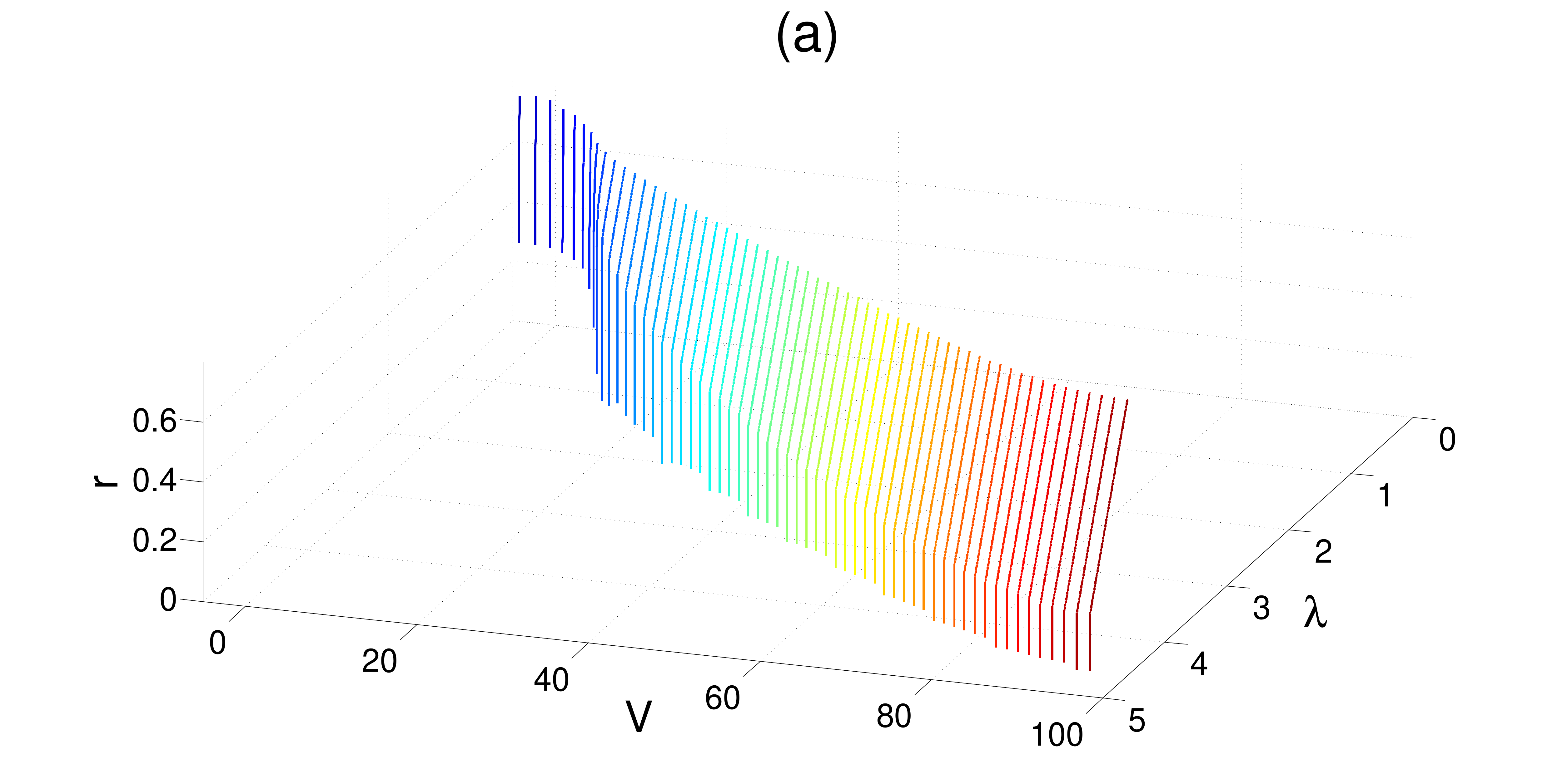}
\par\end{centering}

\noindent \begin{centering}
\includegraphics[scale=0.35]{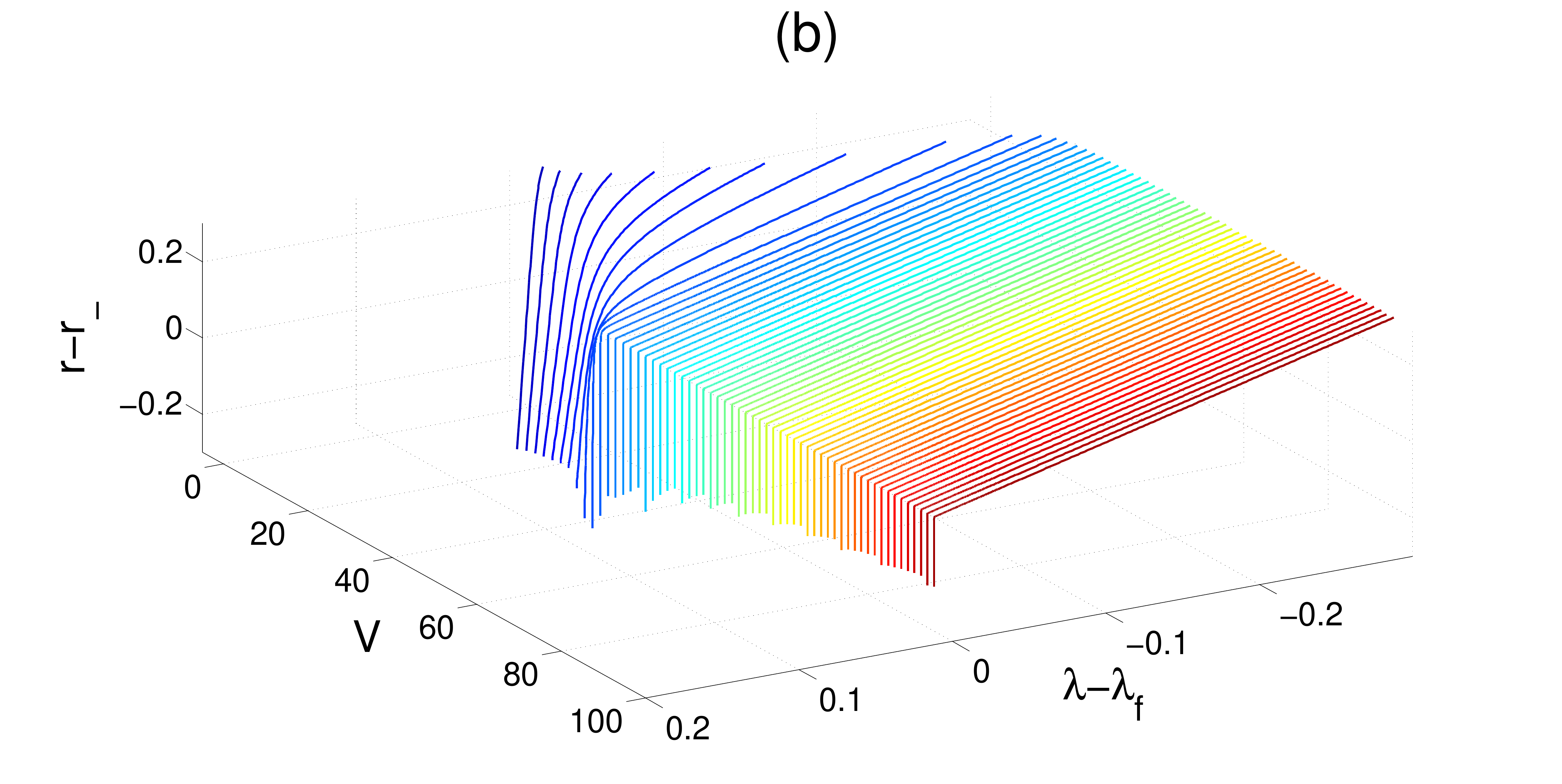}
\par\end{centering}

\protect\caption{\label{fig:DNF-shock-3d} Gravitational shock wave in case of two
null fluids. Panel (a) displays $r(\lambda)$ for a series of ingoing
radial null geodesics; panel (b) displays the same set of geodesics,
shifted in $r$ and $\lambda$ in a similar fashion to Fig. \ref{fig:DNF-shock-2d}(d);
here we subtract the specific $r_{-}(V)$ value of each geodesic from
its $r(\lambda)$ and shift $\lambda$ by $-\lambda_{f}$ (the specific
final $\lambda$ value of the geodesic). The geodesics are solid lines,
distinguished by different colors. The set consists of grid null rays
at the range $-3.3<V<95.7$ with irregular $V$ values and increments
of $V$; the series of Fig. \ref{fig:DNF-shock-2d}(b) is a subset
of this series. In both panels the results are cut in $r$ and $\lambda$
in order to allow visibility. The shock manifests as a sharp ``break''
or drop in $r(\lambda)$ at the IH ($r_{-}(V)$). Panel (b) demonstrates
this better due to the shift.}
\end{figure}

\par\end{center}

\subsection{Shock sharpening rate\label{sub:Shock-SharpRate-DNF}}

MO predicted from analytical considerations that any characteristic
$\Delta\tau$ width of the shock %
\footnote{Characteristic width was (generally) defined as the proper time duration
between two points near the IH on a timelike geodesic on which the
perturbation (or $r)$ receives different values. We follow MO and
EO and focus our sharpening rate analysis on timelike geodesics as
well. %
} is expected to decrease exponentially according to $\Delta\tau\sim e^{-\kappa_{-}v_{eh}}$,
where $\kappa_{-}$ was the constant RN (or Kerr) surface gravity
at the IH and $v_{eh}$ was their timing parameter for timelike geodesic
--- the value of Eddington advanced time coordinate at the EH. EO
have confirmed MO prediction numerically. We do not expect this relation
to hold in our case since we have a slowly growing surface gravity
at the IH. We also have different timing parameters, the aforementioned
$V_{h}$ and $\widetilde{V}_{f}$ (defined at Sec. \ref{sub:T-parameter-def}
and as yet unused). We expect that if we chose our timing parameters
wisely they should admit a simple analytical expression for the shock
sharpening rate. 

In order to analyze the sharpening rate, we first need to define a
specific characteristic $\Delta\tau$ width of the shock. We again
follow EO ansatz and define $\Delta\tau$ as the proper time duration
to drop from $0.75r_{-}(v)$ to $0.25r_{-}(v)$ along the geodesic.
\footnote{Note, however, that in EO case $r_{-}$ was constant (the IH value
of the asymptotic RN BH) and in our case it is a function of $v$.%
} Fig. \ref{fig:DNF-Shock-Width} presents the rapid decrease in this
characteristic width as a function of the timing parameter $V_{h}$.
Each point in this figure represents the characteristic shock width
of a single timelike geodesic from Fig. \ref{fig:DNF-shock-2d}(a);
our characteristic width is unavailable for geodesics in the range
$V_{h}<25$ since they do not reach $r=0.25r_{-}(v)$ in our simulation.
The decrease in $\ln(\Delta\tau)$ is very rapid, reaching extreme
orders of $\ln(\Delta\tau)\sim-1200$; the non-linear pattern of the
decrease confirms our suspicion that the simple exponential law of
MO is no longer valid. 

We consider two new generalized sharpening rate laws in Fig. \ref{fig:DNF-Sharp-Rate},
each associated with a different timing parameter. We consider the
possible sharpening rate law  $\Delta\tau\sim e^{-\intop\kappa_{-}(V_{h})\, dV_{h}}$
in panel (a), by comparing $\kappa_{-}(V_{h})$ to the derivative
$-\frac{d\ln(\Delta\tau)}{dV_{h}}$; we consider the possible law
$\Delta\tau\sim e^{-\intop\kappa_{-}(\widetilde{V}_{f})\, d\widetilde{V}_{f}}$
in panel (b) by comparing $\kappa_{-}(\widetilde{V}_{f})$ to the
derivative $-\frac{d\ln(\Delta\tau)}{d\widetilde{V}_{f}}$. %
\footnote{The associated $\kappa_{-}$ value of each geodesic is calculated
by submitting $m_{BH}(v)$ at the last $v$ value of the geodesic
(deep inside the shock) in Eq. (\ref{eq:kappa_plus_minus}). We then
associate this value with the chosen timing parameter of the geodesic,
$V_{h}$ or $\widetilde{V_{f}}$. %
} The derivatives and $\kappa_{-}$ curves were calculated using Matlab
spline cubic interpolation. Both sharpening rate laws offers a match
that seems too good to be coincidental; we argue that the second sharpening
rate law  offers a slightly better match due to a better asymptotic
match. 

\begin{center}
\begin{figure}[H]
\begin{centering}
\includegraphics[scale=0.35]{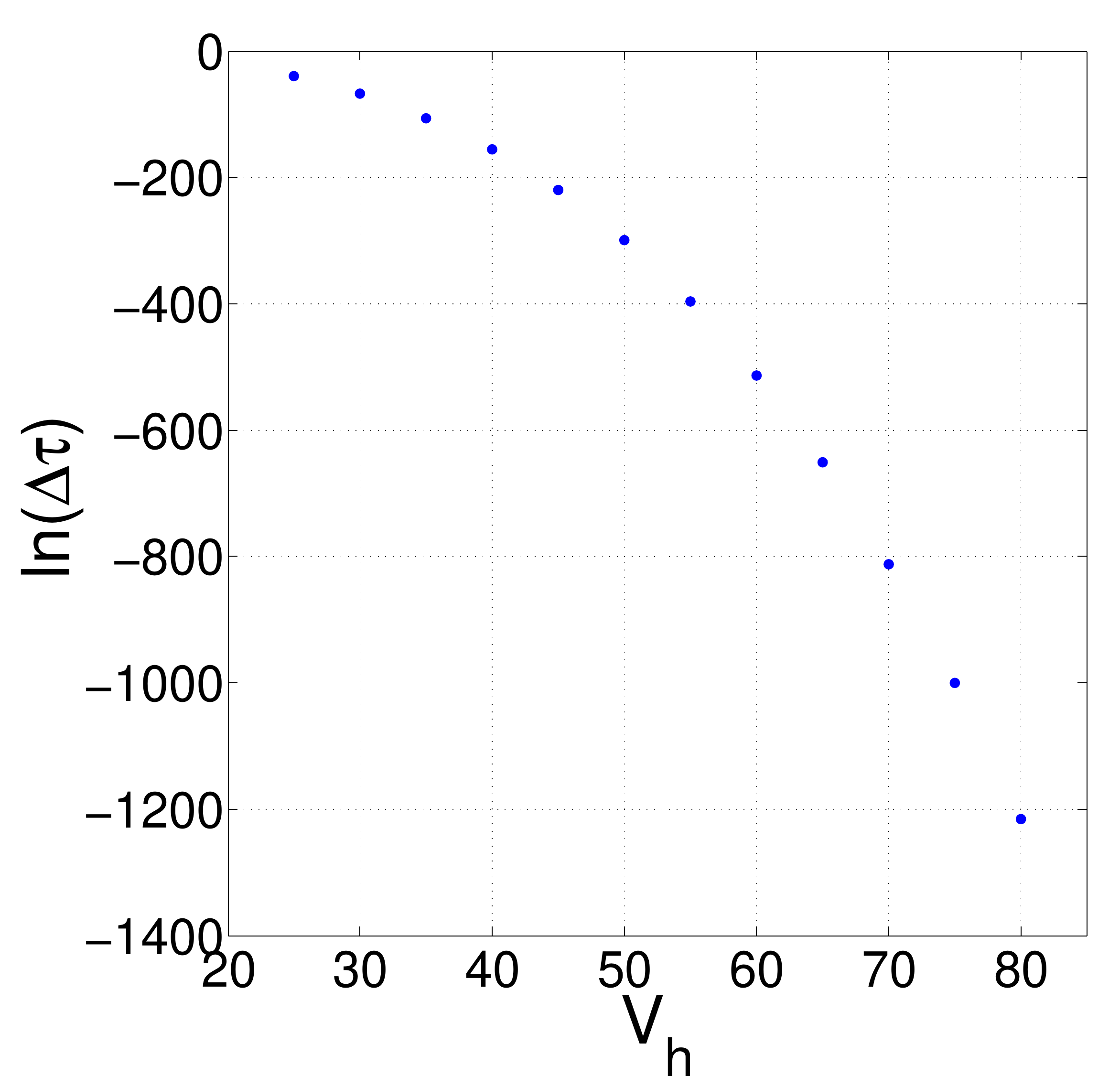}
\par\end{centering}

\protect\caption{\label{fig:DNF-Shock-Width} Gravitational shock width decrease in
the case of two null fluids. The graphs presents $\ln(\Delta\tau)$
as a function of the timing parameter $V_{h}$. Each point represents
a single radial timelike geodesic from the family of Fig. \ref{fig:DNF-shock-2d}(a)
(at the range $25\leq V_{h}\leq80$, where $V_{h}$ increases by increments
of $5$). The nonlinear behaviour of $\ln(\Delta\tau)$ suggests that
the original MO's linear relation indeed breaks as expected due to
the change in $\kappa_{-}(V_{h}).$ }
\end{figure}

\par\end{center}

\begin{center}
\begin{figure}[H]
\begin{centering}
\includegraphics[scale=0.35]{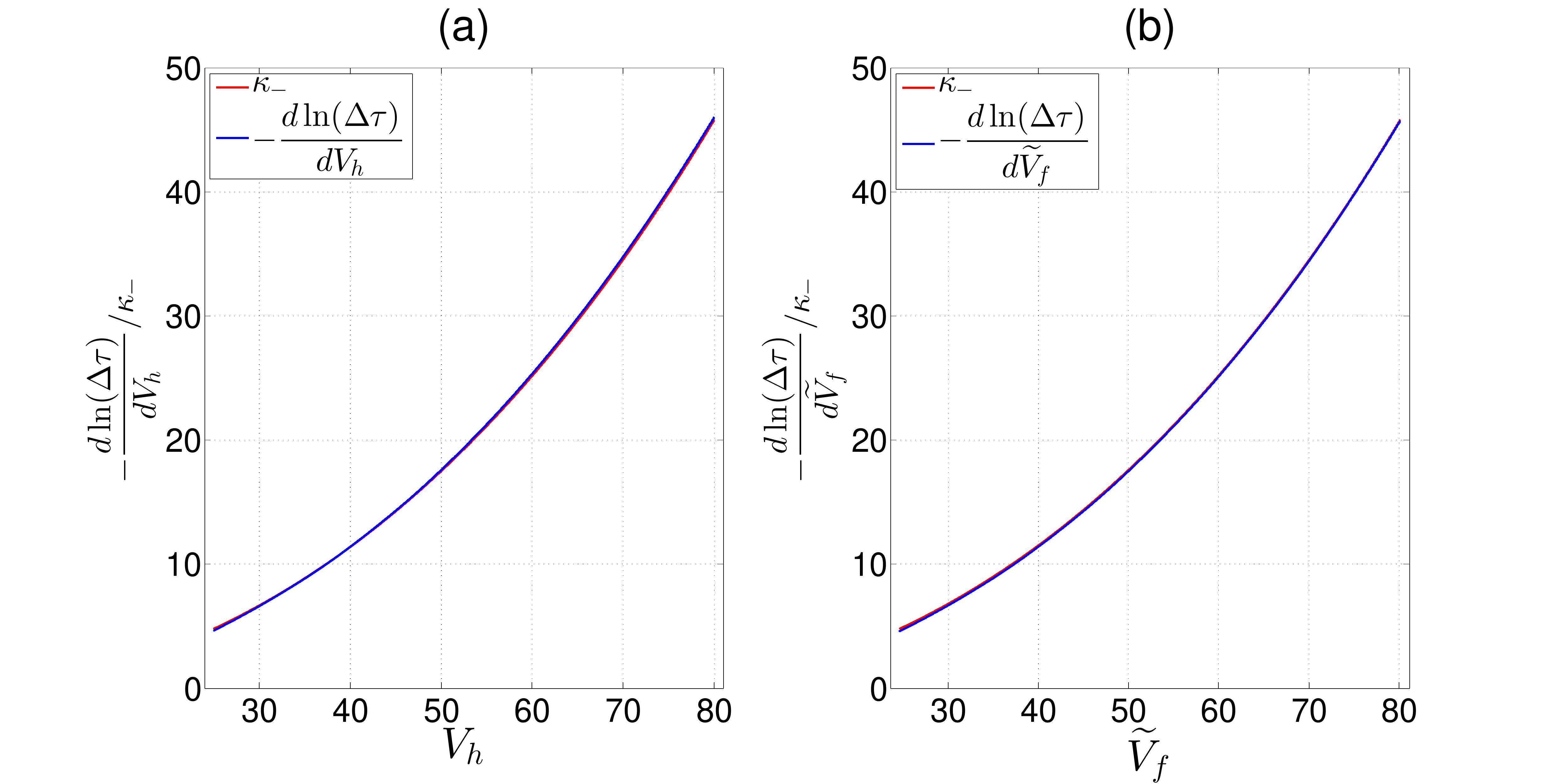}
\par\end{centering}

\protect\caption{\label{fig:DNF-Sharp-Rate} Shock sharpening rate in the case of two
null fluids. Panel (a) tests the possible connection $\Delta\tau\sim e^{-\intop\kappa_{-}(V_{h})\, dV_{h}}$
by comparing the increasing $\kappa_{-}(V_{h})$ to the derivative
$-\frac{d\ln(\Delta\tau)}{dV_{h}}$; panel (b) tests the possible
connection $\Delta\tau\sim e^{-\intop\kappa_{-}(\widetilde{V}_{f})\, d\widetilde{V}_{f}}$
by comparing $\kappa_{-}(\widetilde{V}_{f})$ to the derivative $-\frac{d\ln(\Delta\tau)}{d\widetilde{V}_{f}}$.
Both relations provide a fairly good match; we argue there is a slight
advantage for the second ($\Delta\tau\sim e^{-\intop\kappa_{-}(\widetilde{V}_{f})\, d\widetilde{V}_{f}}$
of panel (b)), due to a better asymptotic match. The calculations
of the derivatives and $\kappa_{-}$ curves are based on Matlab cubic
spline interpolation of data from the geodesics of Fig. \ref{fig:DNF-Shock-Width}.}
\end{figure}

\par\end{center}

\section{THE MIXED CASE: AN INGOING NULL FLUID WITH A SELF GRAVITATING SCALAR
FIELD \label{sec:mixed}}

We construct our mixed case through the addition of an ingoing self-gravitating
scalar-field pulse to the initial data setup described in Sec. \ref{sub:IC-RNV}.
We also slightly modify our ingoing null-fluid stream in order to
preserve its mass contribution. EO have already demonstrated the existence
of a gravitational shock in the self-gravitating scalar-field case;
here we are interested in the effect of the null-fluid stream on a
previously generated shock. Again, we demonstrate the shock presence
and analyze its sharpening rate by the same methods of the previous
section.

\subsection{Basic parameters and initial conditions \label{sub:IC-Mixed}}

The initial RN spacetime has the same parameters as in the previous
two cases --- initial mass parameter of $M_{0}=1$ and charge parameter
of $Q=0.95$. We define an ingoing self-gravitating scalar-field pulse
on the initial ray $u=u_{0}$ in the same fashion as EO did, 

\begin{center}
\begin{equation}
\Phi(u_{0},v)=\begin{cases}
\begin{array}{c}
A_{3}\frac{64(v-v_{1})^{3}(v_{2}-v)^{3}}{(v_{2}-v_{1})^{6}}\\
0
\end{array} & \begin{array}{c}
|\,\,\, v_{1}\leq v\leq v_{2}\\
|\,\,\, otherwise\,,
\end{array}\end{cases}\label{eq:SF_V}
\end{equation}

\par\end{center}

\noindent where $A_{3}$ is an amplitude parameter. Similar to the
outgoing null-fluid pulse from the previous section, this scalar-field
pulse has compact support on the initial ray $u=u_{0}$, and is limited
to a certain $v$ range, $v_{1}\leq v\leq v_{2}$. We choose $v_{1}=1$
and $v_{2}=7$ (like EO did) %
\footnote{This range is equivalent to $-6.64\leq V\leq0$.%
} but a slightly higher value of $A_{3}$ than EO's ($A_{3}=0.1201$)
in order to set the mass contribution of the pulse to the BH to be
$\Delta m_{\Phi,vmax}\simeq0.5$. The relative location of the pulse
and its shape are illustrated in panel (a) of Fig. \ref{fig:Penrose-Mixed}.
The ingoing null fluid stream has the same form of Eq. (\ref{eq:Tvv_shape}),
but we slightly raise its amplitude as well (to $A_{1}=4.267591\times10^{-4}$)
in order to preserve the mass contribution to the BH ($\Delta m_{NF,vmax}\simeq1.5$).
The stream begins at $r_{1}=10$, which is now equivalent to $v_{3}\simeq14.77$
or $V_{3}\simeq7.35$. Overall we get a BH with a mass of $m_{BH,vmax}\simeq3.0$;
the corresponding EH and IH values are $r_{+,vmax}\simeq5.846$ and
$r_{-,vmax}\simeq0.1544$, and the corresponding IH surface gravity
is $\kappa_{-,vmax}\simeq119.4$. The EH is reached earlier than the
previous two cases, at $u_{h}\simeq42.9$. 

The remaining initial values are taken according to the third column
of table \ref{tab:IC-def}. In particular, $T_{vv}^{NF}$ vanishes
on $v=v_{0}$; $T_{uu}^{NF}$ vanishes on both initial rays; the scalar
field $\Phi$ vanishes on $v=v_{0}$ and $\sigma$ conforms on both
rays with the maximal-$\sigma$ gauge condition (Eq. (\ref{eq: Sigma_max:u+v})).

The domain of integration is $u_{0}=v_{0}=0$, $u_{max}=298.65$,
and $v_{max}=120$. The value of $r$ in the initial vertex is $r_{0}=5$
and it grows up to $r\simeq40.02$ at $r(u_{0},v_{max})$. The third
corner of the grid is $r(u_{max},v_{0})\simeq0.1003$; the fourth
corner $(u_{max},v_{max})$ is located (again) beyond a spacelike
singularity.

\subsection{The structure of spacetime\label{sub:ST-Mixed}}

The structure of spacetime in this case (displayed on Fig. \ref{fig:Penrose-Mixed})
is simpler than the case of two null fluids in several respects: there
are only four patches of spacetime, and only three of them are covered
in our simulation; the borders between the patches are always $v=const$
rays; spacetime is (mostly) non extendable due to presence of a (well
known) spacelike $r=0$ singularity. %
\footnote{The ingoing CH at the border of the last patch PC3 also contains a
well known curvature singularity, though a weak one. While this singularity
does not prevent the extension of spacetime in the physical sense,
the extension is not well defined (not unique).%
} The initial RN patch ends at $v=v_{1}$; it is followed by perturbed
charged patch (PC1) which contains a self gravitating scalar-field
perturbation. The spacelike singularity and the gravitational shock
wave originate from this patch; it ends at the beginning of the null
fluid stream at $v=v_{3}$. The second perturbed charged patch (PC2)
contains the null fluid stream as well as the (slowly decaying) scalar-field;
this is the most interesting patch to us, and it ends with the null
fluid stream at $v=v_{4}$. The last perturbed charged patch (PC3)
contains, again, just the slowly decaying scalar field. Asymptotically,
the geometry of this patch outside the BH and at the close vicinity
of the EH is RN geometry with some unknown mass $m_{final}$. This
patch is not covered in our simulation. 

The numerical results for $r(u,v)$ (Fig. \ref{fig:Mixed-ruv}) demonstrate
the spacelike $r=0$ singularity very clearly; this time it is even
noticeable at the full grid level (panel (a)), although a closer zoom
(panel (b)) is needed in order to reveal its spacelike nature (the
curved shape of the border of the criss cross patch). This zoom also
offers us some puzzlement in the form of two seemingly null sections
of the singularity; one vertical $v=const$ section of the border
(begins at $u\sim295$) and one horizontal $u=const$ section of the
border (begins at $v\sim7$). We focus on these sections on panels
(c) and (d). We argue that these sections are not likely to represent
true null singularity sections, however, for several reasons. Panels
(c) and (d) reveal a smooth transition between these sections and
spacelike sections; the truly null sections appear much shorter in
these panels. %
\footnote{The vertical null section seems to begin at $u\sim295$ on panel (b)
and on $u\sim298$ on panel (c); the horizontal null section seems
to begin at $v\sim11$ on panel (b) and on $v\sim12.8$ on panel (d).
Further zoom on panel (c) shortens the vertical section even further;
further zoom on panel (d), however, does not change the length of
the horizontal section, perhaps due to resolution limit. %
} The quick succession of null-spacelike-null-spacelike sections raises
further skepticism. Lastly, these sections are located mostly in the
patch PC1, before the null fluid stream (the vertical section is entirely
in PC1; the horizontal section ends at $v\sim15$, where PC2 begins
at $v\simeq14.77$); scalar-field perturbation is not known to generate
null $r=0$ singularities. We tend to conclude that these sections
are actually spacelike sections and suggest that the changes in appearance
of the spacelike singularity represent changes of phases in spacetime
{[}e.g. the end of the main scalar field pulse (on $v=7$ at $u=u_{0}$,
slightly later on higher $u$ values), and the beginning of the null
fluid stream{]}. We are aware this explanation raises further doubts
regarding the nature of the suspected null $r=0$ singularity in the
two null fluids case (Sec. \ref{sub:ST-DNF}); we have already agreed
that further research is needed in order to confirm its existence.

\begin{center}
\begin{figure}[H]
\noindent \begin{centering}
\includegraphics[scale=0.5]{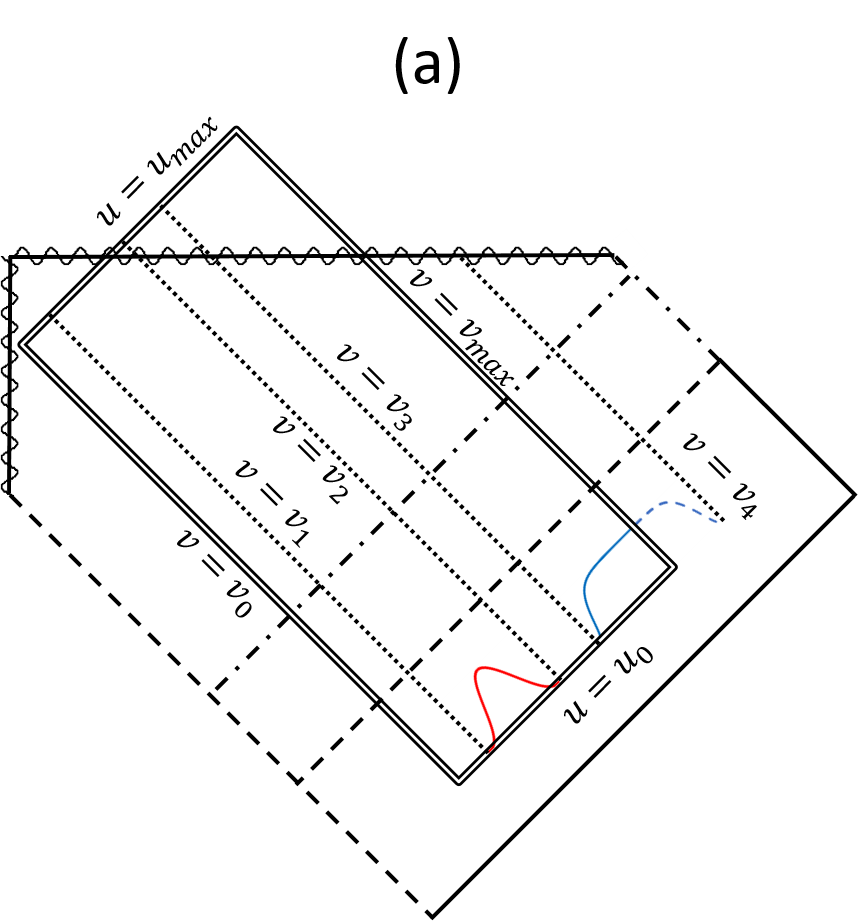}\hspace{2cm}\includegraphics[scale=0.5]{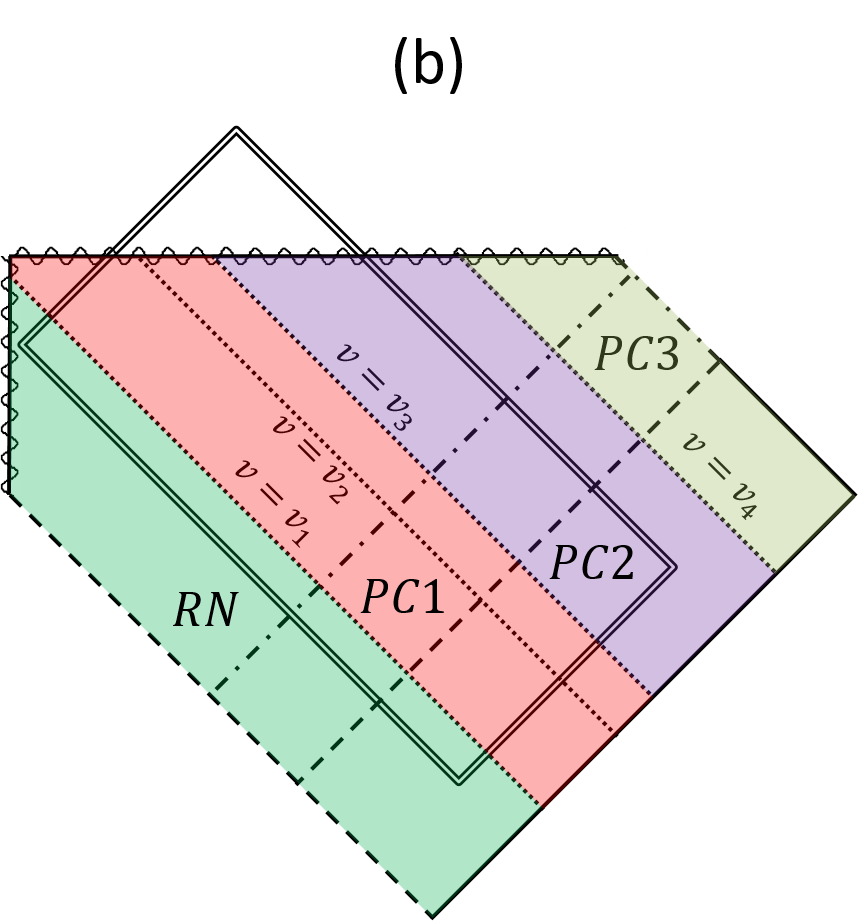}
\par\end{centering}

\protect\caption{\label{fig:Penrose-Mixed} Penrose diagrams illustrating the structure
of spacetime in the mixed case. Both panels describe the location
of the numerical domain of integration and the different perturbations
in spacetime; while panel (a) focuses on initial data, panel (b) analyzes
the different patches of spacetime. In both panels, the domain's limits
are denoted by a double black line; solid black lines denote null
infinity; dashed black lines denote the EH; dashed-dotted black lines
denote the IH. Wavy lines denote the timelike $r=0$ singularity of
the initial RN geometry, as well as the spacelike $r=0$ singularity
that develops in this case. Dotted black lines denote the limits of
the ingoing scalar-field pulse on the initial ray $u=u_{0}$ ($v=v_{1}$
and $v=v_{2}$) and the ingoing null-fluid stream limits $v=v_{3}$
and $v=v_{4}$. The blue curve in panel (a) represents the ingoing
null-fluid stream on the initial ray $u=u_{0}$. The red curve represents
the ingoing scalar-field pulse on the same initial ray; its shape
reflects directly the shape of the pulse in $\Phi$. Panel (b) demonstrates
that spacetime could be divided into four distinct patches: one patch
of initial RN geometry (RN), and three patches of perturbed charged
geometry (PC1, PC2 and PC3). PC1 and PC3 are perturbed only by the
scalar field; PC2 is perturbed by the null fluid stream and the scalar
field. The patch PC3 is not covered in our simulation and has asymptotically
RN geometry outside the BH (and at the close vicinity of the EH). }
\end{figure}

\par\end{center}

\begin{center}
\begin{figure}[H]
\noindent \begin{centering}
\includegraphics[scale=0.35]{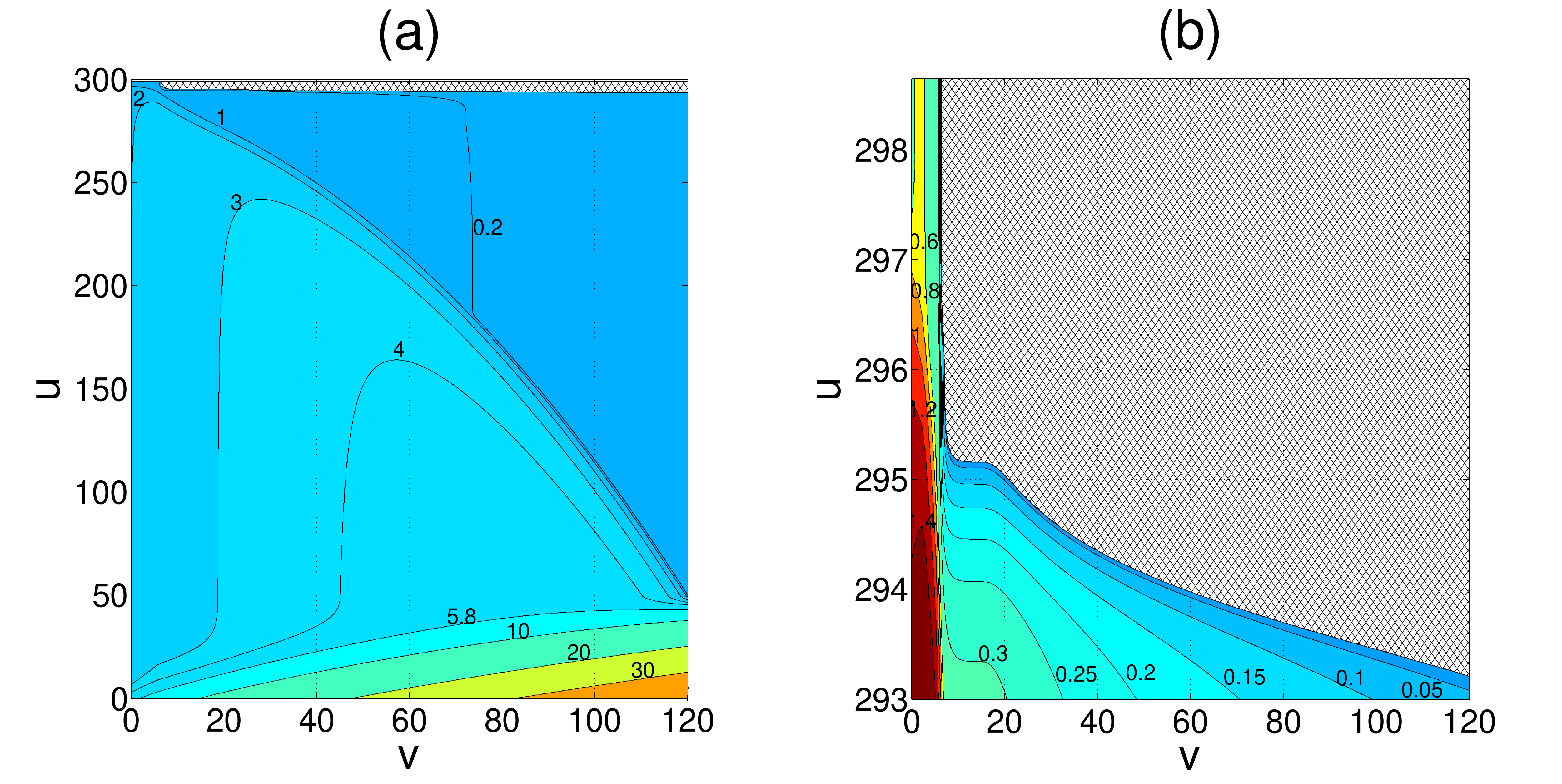}
\par\end{centering}

\noindent \begin{centering}
\includegraphics[scale=0.35]{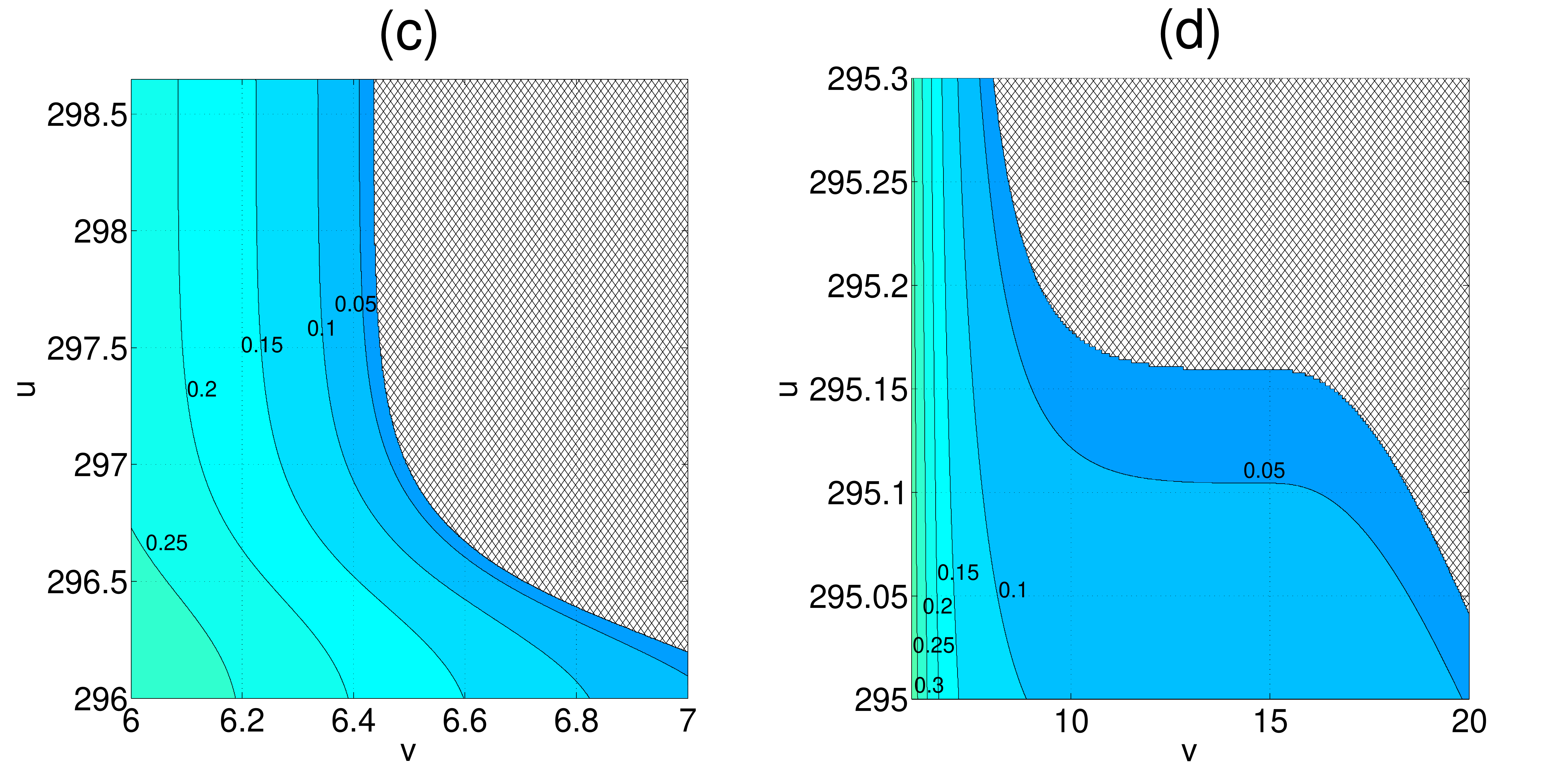}
\par\end{centering}

\protect\caption{\label{fig:Mixed-ruv} Numerical results for $r(u,v)$ in the mixed
case. The panels display contour graphs of $r(u,v)$ based on numerical
results; panel (a) displays results on the entire grid; panel (b)
displays a zoom deep inside the BH, near $u=u_{max}=298.65$; panels
(c) and (d) display further zooms on panel (b). Panel (a) uses different
color code and different level choice for $r$ from panels (b)-(d)
for the sake of visibility. The criss-cross patch in all panels represents
a region in which numerical results are unavailable due to the $r=0$
singularity. Panel (b) reveals the presence of a spacelike $r=0$
singularity; it manifests as the curved border of the criss cross
patch. There are also a horizontal section and a vertical section
of the border in panel (b) that could have hinted on the existence
of two sections of null singularity; we argue in the text that this
is not likely here. Panels (c) and (d) focus on these sections. Panel
(a) is based on $N=640$ numerical results sampled in a lower resolution
$(\Delta_{s}u=\Delta_{s}v=0.1)$; panel (b) is based on $N=640$ results
with a sampling on $v$ alone $(\Delta_{s}v=0.05)$; panels (c) and
(d) are based on $N=640$ results without any sampling. }
\end{figure}

\par\end{center}

\subsection{Gravitational shock detection \label{sub:Shock-Mixed}}

The gravitational shock wave is demonstrated in Figs. \ref{fig:Mixed-shock-2d}
and \ref{fig:Mixed-shock-3d}. Panels (a) and (c) of Fig. \ref{fig:Mixed-shock-2d}
display the shock in $r(\tau)$ of timelike geodesics; panels (b)
and (d) display the shock in $r(\lambda)$ of null geodesics. Again,
panels (c) and (d) display the same geodesics as panels (a) and (b)
but with a shift in $r$ and $\tau/\lambda$ in order to allow clearer
observation of the shock. The general picture is very similar to the
case of two null fluids; the shock manifests as a clear vertical drop
in $r(\tau)/r(\lambda)$, located at the appropriate $r_{-}$ value
of the geodesic. The main difference is that now the shock begins
earlier, due to the scalar-field pulse; it could be observed as early
as the timelike geodesic $V_{h}=5$ or the null geodesic $V=8.4$.
In panels (a) and (b), one could notice the short break between the
end of the scalar field pulse and the beginning of the null fluid
stream on which $r_{-}$ is roughly constant; this break is located
between the timelike geodesics $V_{h}=0$ and $V_{h}=5$ in panel
(a) and the null geodesics $V=1$ and $V=8.4$ in panel (b). %
\footnote{Although the shock is not well developed in the geodesics $V_{h}=0$
and $V=1$ and $V=8.4$ is already inside the null fluid stream.%
} After this break $r_{-}(v)$ continues to decrease.

We demonstrate the ``vertical wall'' representation of the shock
through null geodesics in a three-dimensional graph in Fig. \ref{fig:Mixed-shock-3d}.
Panel (a) displays the true $r(\lambda)$ curves of the geodesics
while panel (b) displays shifted curves. Again, we use geodesics from
the same range of Fig. \ref{fig:Mixed-shock-2d}(b) but with a denser
sampling. Here, unlike the case of two null fluids, all the geodesics
seem ``broken'', each one breaks to a sheer drop at a different
$r$ value. Panel (b) confirms that this is indeed the right $r_{-}(V)$
value. 

\begin{center}
\begin{figure}[H]
\noindent \begin{centering}
\includegraphics[scale=0.35]{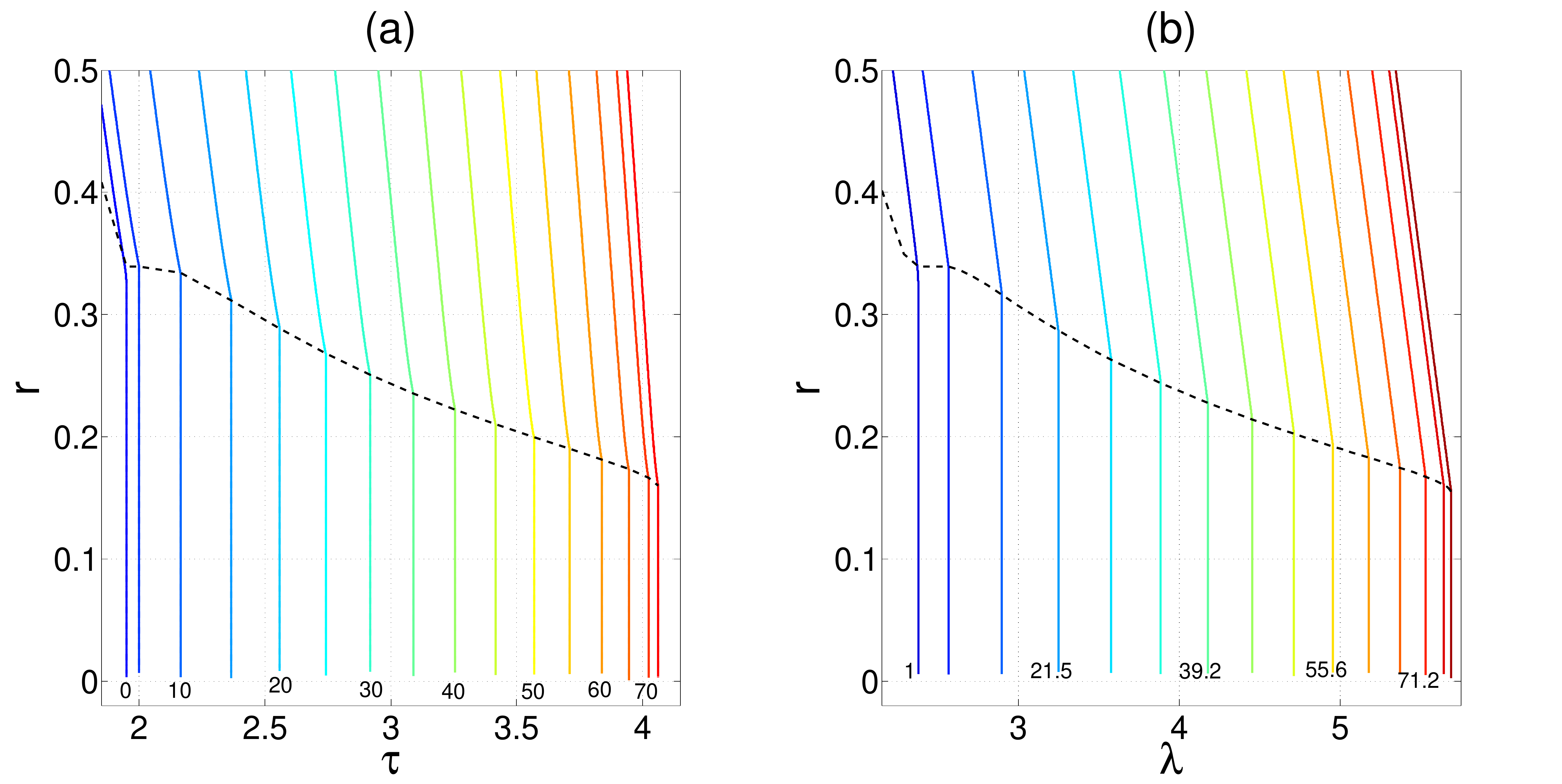}
\par\end{centering}

\noindent \begin{centering}
\includegraphics[scale=0.35]{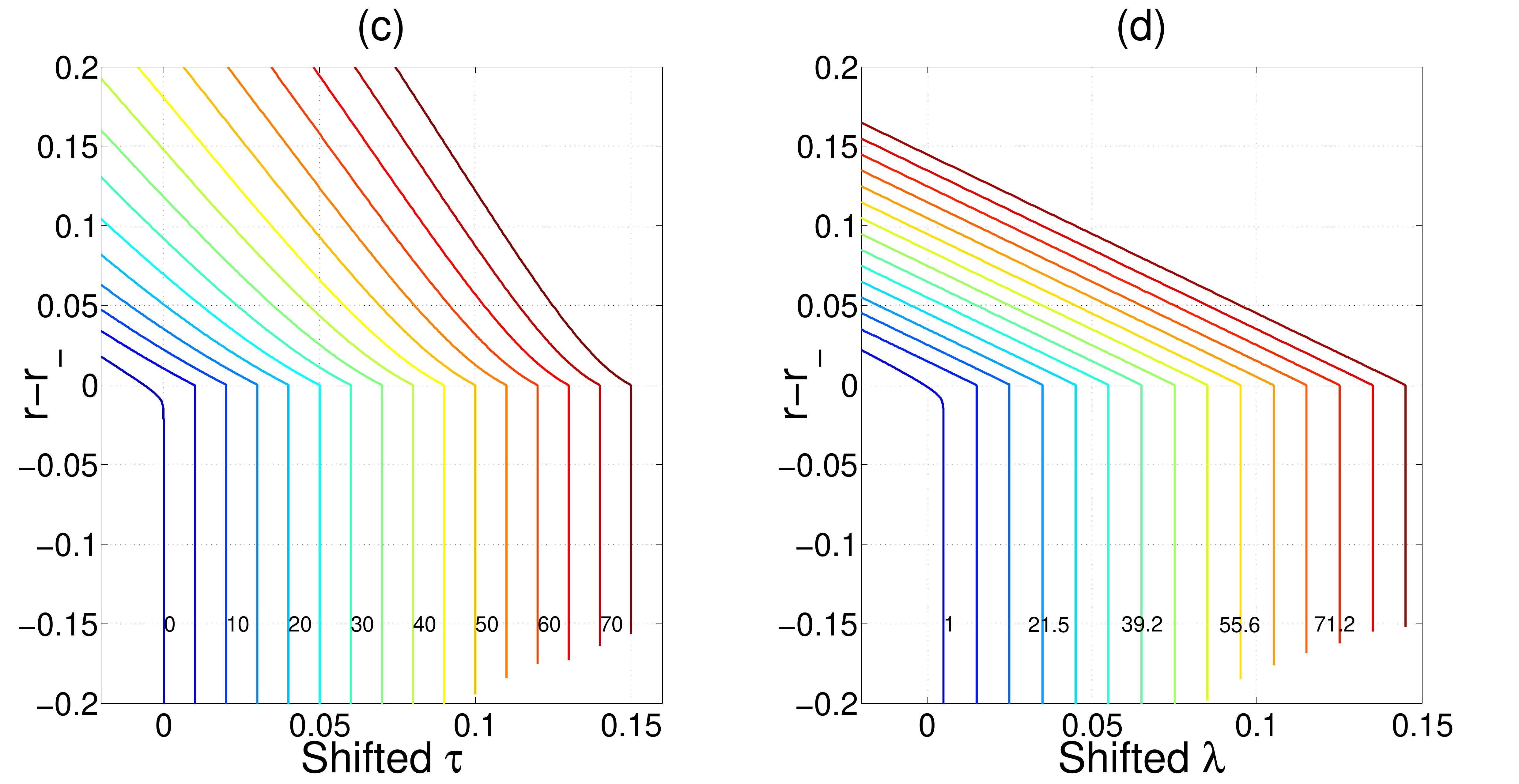}
\par\end{centering}

\protect\caption{\label{fig:Mixed-shock-2d} Gravitational shock wave in the mixed
case. Panel (a) presents $r(\tau)$ for a family of radial timelike
geodesics; panel (b) presents $r(\lambda)$ for a series of ingoing
radial null geodesics. Geodesics are solid lines, distinguished by
different colors and different timing parameter values; the timing
parameter is $V_{h}$ for timelike geodesics and $V$ for null geodesics.
The dashed black curve in panels (a) and (b) denotes the evolving
IH value of the geodesics; $r_{-}(V_{h})$ in panel (a) and $r_{-}(V)$
in panel (b). Panels (c) and (d) display the same set of geodesics
as panels (a) and (b) (accordingly) but shifted by constant factors
in the same fashion as panels (c) and (d) of Fig. \ref{fig:DNF-shock-2d}.
Panels (a) and (c) contain timelike geodesics at the range $0\leq V_{h}\leq75$,
where $V_{h}$ increases by increments of $5$ from left $(V_{h}=0$)
to right $(V_{h}=75)$; panels (b) and (d) contain grid null rays
at the range $0<V<81.5$, with irregular $V$ values and irregular
increments of $V$ from left $(V\simeq1)$ to right $(V\simeq81.4)$.
Gravitational shock manifests as a sharp drop or ``break'' in $r(\tau)/r(\lambda)$
at the IH; the shock is first seen clearly on $V_{h}=5/V=8.4$ (the
second geodesic from the left in panels (a) and (c)/panels (b) and
(d)), and is easier to observe in panels (c) and (d).}
\end{figure}

\par\end{center}

\begin{center}
\begin{figure}[H]
\noindent \begin{centering}
\includegraphics[scale=0.35]{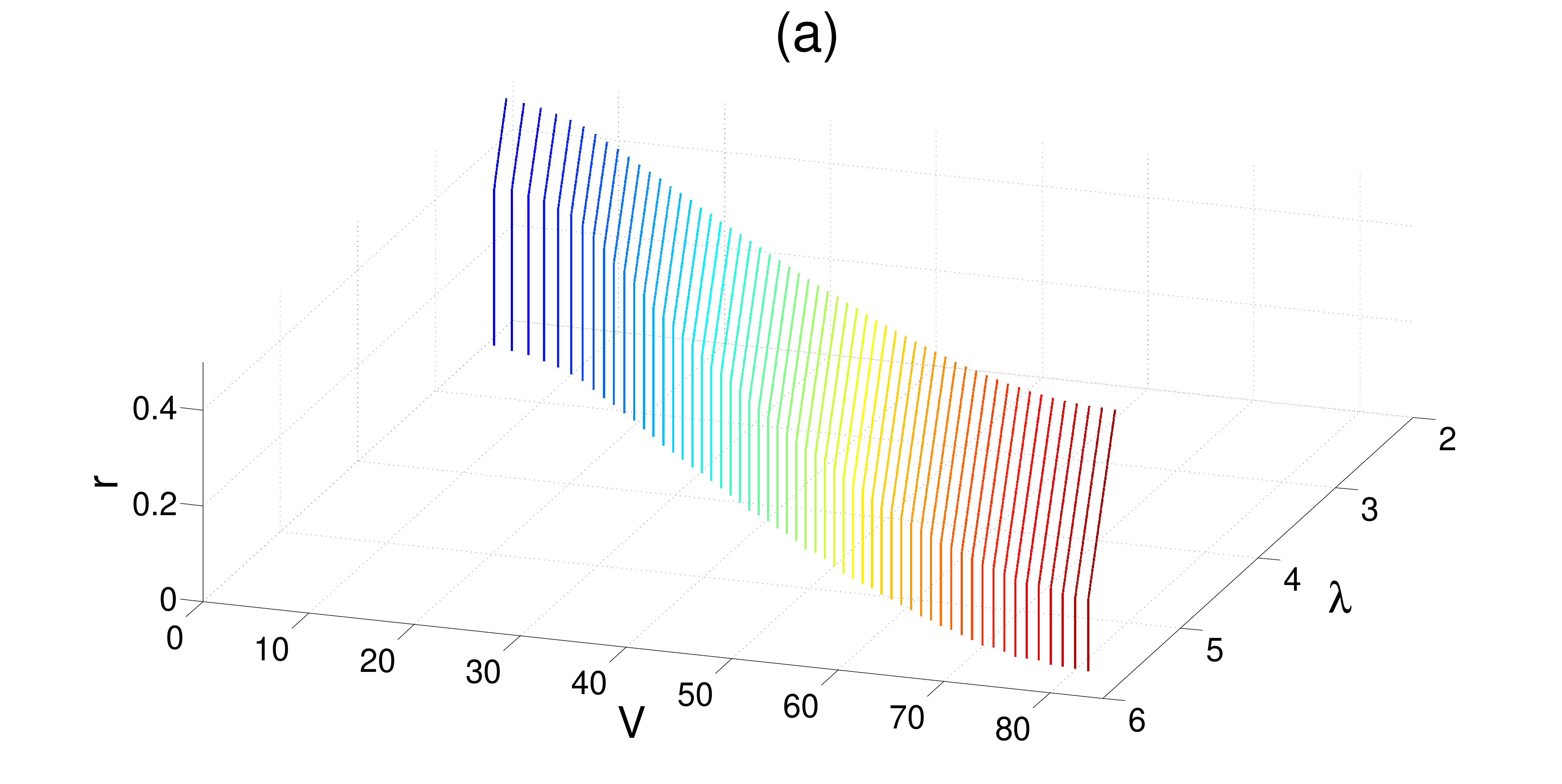}
\par\end{centering}

\noindent \begin{centering}
\includegraphics[scale=0.35]{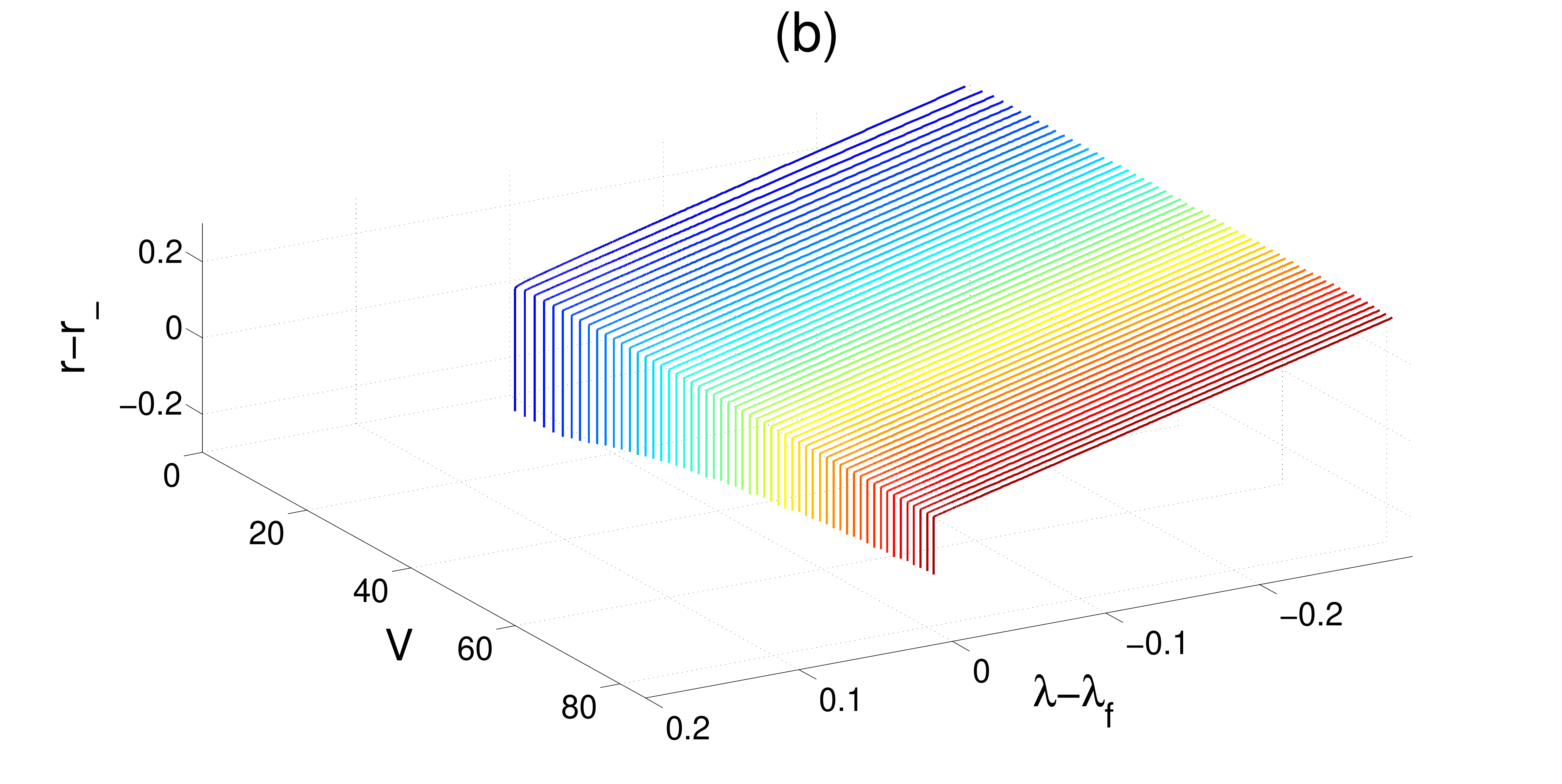}
\par\end{centering}

\protect\caption{\label{fig:Mixed-shock-3d} Gravitational shock wave in the mixed
case. Panel (a) displays $r(\lambda)$ for a series of ingoing radial
null geodesics; panel (b) displays the same set of geodesics, shifted
in $r$ and $\lambda$ in a similar fashion to Fig. \ref{fig:DNF-shock-3d}(b).
The geodesics are solid lines, distinguished by different colors.
The set consists of grid null rays at the range $0<V<81.5$ with irregular
$V$ values and increments of $V$; the series of Fig. \ref{fig:Mixed-shock-2d}(b)
is a subset of this series. In both panels the results are cut in
$r$ and $\lambda$ in order to allow visibility. The shock manifests
as a sharp ``break'' or drop in $r(\lambda)$ at the IH ($r_{-}(V)$).
Panel (b) demonstrates this better due to the shift.}
\end{figure}

\par\end{center}

\subsection{Shock sharpening rate\label{sub:Shock-SharpRate-Mixed}}

We turn next to analyze the sharpening rate of the shock by the same
method used in Sec. \ref{sub:Shock-SharpRate-DNF}. Fig. \ref{fig:Mixed-Shock-Width}
presents the decrease in the characteristic width of the shock $\Delta\tau$
(the same width defined in Sec. \ref{sub:Shock-SharpRate-DNF}). Each
point represents a single timelike geodesic from Fig. \ref{fig:Mixed-shock-2d}(a);
in this case we have characteristic widths for all of them. The decrease
in $\ln(\Delta\tau)$ is extreme, reaching orders of $\ln(\Delta\tau)\sim-3250$.
Fig. \ref{fig:Mixed-Sharp-Rate} considers the same two suggested
sharpening rate laws as Fig. \ref{fig:DNF-Sharp-Rate}, $\Delta\tau\sim e^{-\intop\kappa_{-}(V_{h})\, dV_{h}}$
in panel (a) and $\Delta\tau\sim e^{-\intop\kappa_{-}(\widetilde{V}_{f})\, d\widetilde{V}_{f}}$
in panel (b), by checking the relevant matches between $\kappa_{-}$
and the derivative of $-\ln(\Delta\tau)$. The match for both laws
is, again, fairly good but not perfect; we notice a clear difference
between the curves of $\kappa_{-}$ and the derivatives in the early
phase ($V_{h}\apprle30,\widetilde{V}_{f}\lesssim45$) as well as additional
difference for the first law at interim $V_{h}$ values $(40\apprle V_{h}\apprle70)$.
We (again) argue that the second connection ($\Delta\tau\sim e^{-\intop\kappa_{-}(\widetilde{V}_{f})\, d\widetilde{V}_{f}}$)
provides a better description of the sharpening rate due to the better
match between $\kappa_{-}$ and the relevant derivative of $-\ln(\Delta\tau)$.

\begin{center}
\begin{figure}[H]
\begin{centering}
\includegraphics[scale=0.35]{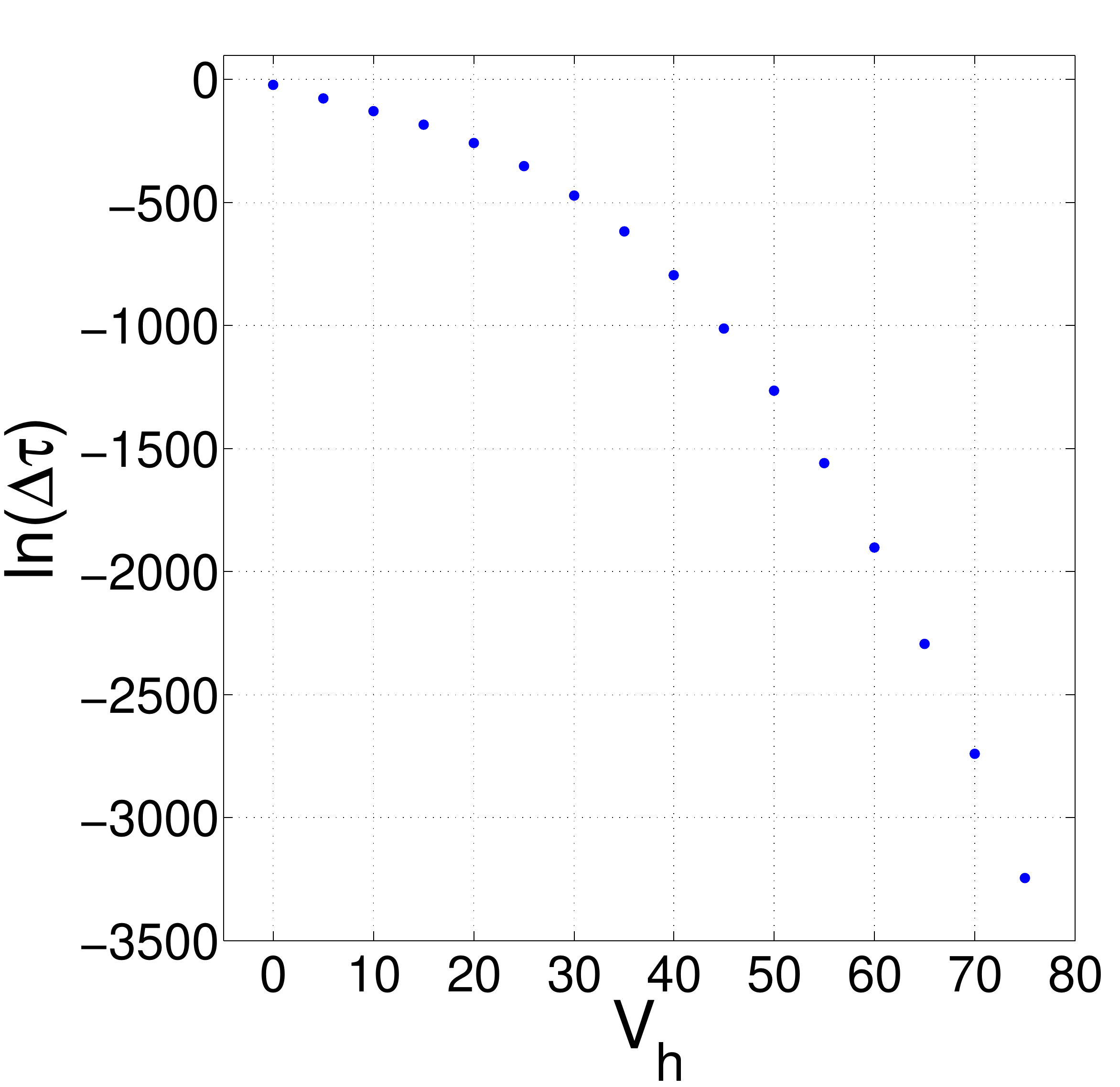}
\par\end{centering}

\protect\caption{\label{fig:Mixed-Shock-Width} Gravitational shock width decrease
in the mixed case. The graphs presents $\ln(\Delta\tau)$ as a function
of the timing parameter $V_{h}$. Each point represents a single radial
timelike geodesic from the family of Fig. \ref{fig:Mixed-shock-2d}(a)
(at the range $0\leq V_{h}\leq75$, where $V_{h}$ increases by increments
of $5$).}
\end{figure}

\par\end{center}

\begin{center}
\begin{figure}[H]
\begin{centering}
\includegraphics[scale=0.35]{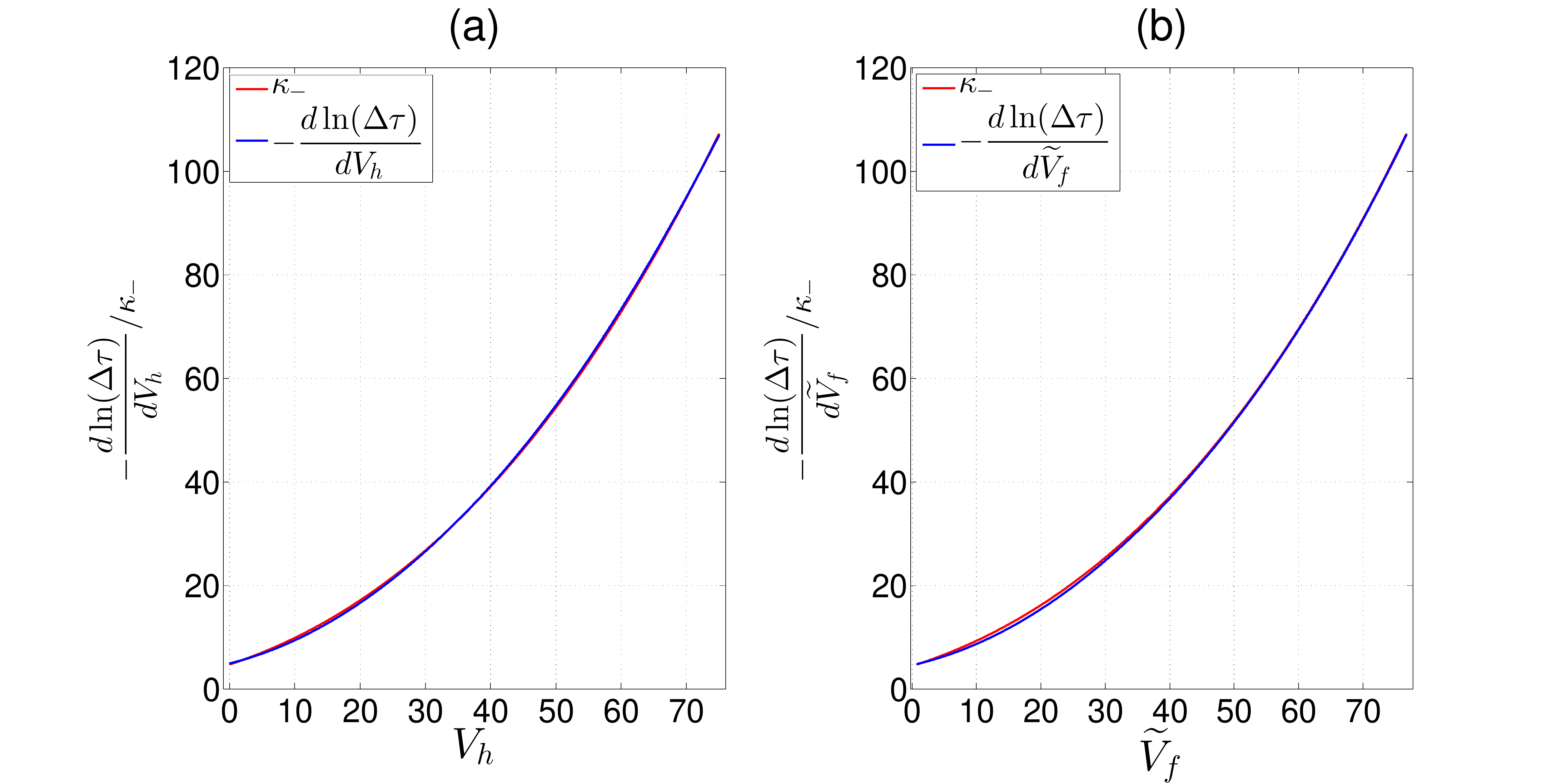}
\par\end{centering}

\protect\caption{\label{fig:Mixed-Sharp-Rate} Shock sharpening rate in the mixed case.
We test the same connections as in Fig. \ref{fig:DNF-Sharp-Rate};
panel (a) tests $\Delta\tau\sim e^{-\intop\kappa_{-}(V_{h})\, dV_{h}}$
and panel (b) tests $\Delta\tau\sim e^{-\intop\kappa_{-}(\widetilde{V}_{f})\, d\widetilde{V}_{f}}$.
Again, both relations provide a fairly good match but the second ($\Delta\tau\sim e^{-\intop\kappa_{-}(\widetilde{V}_{f})\, d\widetilde{V}_{f}}$
of panel (b)), offers a better asymptotic match. The calculations
of the derivatives and $\kappa_{-}$ curves are based on Matlab cubic
spline interpolation of data from the geodesics of Fig. \ref{fig:Mixed-Shock-Width}.}
\end{figure}

\par\end{center}

\section{DISCUSSION \label{sec:discussion}}

Although we did not observe the gravitational shock wave in the single
null fluid case, we have detected it in the case of two null fluids
and in the mixed ingoing null-fluid/self-gravitating scalar-field
case. In both cases the shock manifests as a sharp break or a sheer
drop in $r(\tau)$ (or $r(\lambda)$) of timelike (or null) geodesics,
located at the crossing point of $r=r_{-}(v)$ of the geodesic. $r_{-}(v)$
decreases monotonically due to null fluid accretion; the match between
the shock location to $r=r_{-}(v)$ is conserved throughout the process.
The characteristic width of the shock $\Delta\tau$ decreases rapidly
in both cases. Since MO's original sharpening rate law  is no longer
valid, we have tested two generalized sharpening rate laws through
a match between $\kappa_{-}$ curve and the relevant derivatives of
$-\ln(\Delta\tau)$. We have discovered that the generalized sharpening
rate law $\Delta\tau\sim e^{-\intop\kappa_{-}(\widetilde{V}_{f})\, d\widetilde{V}_{f}}$
offers a fairly good match to the shock sharpening rate in both cases,
though not perfect. In addition, we have gained new insight into the
inner (classical) structure of the BH in the case of two null fluids
perturbation; our numerical $r(u,v)$ results provided strong evidence
for the existence of a spacelike $r=0$ singularity, and a possible
indication for the existence of a null non-naked $r=0$ singularity,
although this indication is uncertain and requires further research.

Our gravitational shock differs from the one EO discussed (in the
self-gravitating scalar field case) in two main respects. Our $r_{-}(v)$
decreases due to long term accretion; EO observed MO original sharpening
rate law of the shock, $\Delta\tau\sim e^{-\kappa_{-}v_{eh}}$ , while
we observed a generalized law ($\Delta\tau\sim e^{-\intop\kappa_{-}(\widetilde{V}_{f})\, d\widetilde{V}_{f}}$),
although it follows immediately that EO sharpening rate law is just
a private case of our generalized sharpening rate in the case of a
constant $\kappa_{-}$, with the proper change in timing parameter
selection. 

The existence of a spacelike $r=0$ singularity in case of two null
fluids perturbation is not entirely surprising. Besides the well known
spacelike singularity of the self-gravitating scalar-field case, \cite{10-Brady-Smith,14-Burko-spacelike}
a spacelike singularity is known to develop in a special two null
fluids case, where the ingoing and the outgoing null fluid fluxes
are equal and time-independent; \cite{Ori-Homogenous-DNF,Page-etal-Homogenous DNF,Dori-Hoogenous-DNF}
in this case spacetime is homogenous and a spacelike singularity develops
from the outset, without a null singularity at the CH at all. In our
two null fluids case, however, the fluxes are unequal (and time dependent);
as far as we are aware of, the existence of a spacelike $r=0$ singularity
in this case has not been demonstrated before in publications.

The minor mismatch in our sharpening rate law  graphs (the differences
between the curve of $\kappa_{-}$ and the curves of the derivatives
of $-\ln(\Delta\tau)$) may be caused by timing parameter choice.
We discuss at some length in Sec. \ref{sub:v-Vaidya-def} of the appendix
the differences between the variants of RNV advanced time; we argue
that $\widetilde{V}$ should be a ``better'' RNV coordinate than
$V$ deep inside the BH (at the shock region) as it is calculated
closer to the shock. (similar argument should also apply on the choice
to evaluate the timing parameter at the shock location instead of
the EH crossing). It follows logically that $\widetilde{V}_{f}$ is
a better timing parameter than $V_{h}$ as it describes the dynamics
of spacetime inside the BH more accurately. But what if the ``right''
timing parameter actually belongs to a third variant of RNV advanced
time coordinate, calculated in the vicinity of the shock instead of
the EH? Such calculation is technically problematic in the current
version of the code, %
\footnote{The calculation is difficult due to the presence of mass inflation
inside the BH and the unavailability of the derivative $r,_{u}$.
As explained in the appendix, either $m$ or $r,_{u}$ is needed for
the calculation of RNV advanced time coordinate in our algorithm.%
} but may resolve some of the differences between the curves. In the
mixed case we have an even better reason to doubt the accuracy of
our timing parameters choice; as discussed in the appendix, at the
early stages of the simulation, when the scalar field is dominant,
the coordinates of our timing parameters are not ``good'' RNV coordinates,
only approximated RNV coordinates; thus we expect them to be less
successful in describing accurately the development of the shock at
this stage. Indeed, the mismatch in the mixed case is very dominant
in the early stage. The mismatch may be, in theory, an artifact of
our cubic spline interpolation procedure; we do not believe this to
be the case since the interpolated series of Figs. \ref{fig:DNF-Shock-Width}
and \ref{fig:Mixed-Shock-Width} seem to describe a reasonably smooth
function sampled at a decent resolution. Moreover, the mismatch is
more obvious at low values of the timing parameter $V_{h}$ and the
function $\ln(\Delta\tau)$ is actually milder there.

We recognize two possible issues that could potentially compromise
the reliability of our numerical results. The first issue is the fact
that our simulation is entirely classical, while we expect Quantum
Gravity to become relevant at the close vicinity of the $r=0$ singularity.
The second issue is the decline in performance of the numerical solver
near this singularity (discussed in Sec. \ref{sec:Numerical-Algorithm}).
We argue that while Quantum Gravity may replace the $r=0$ singularity
with a regular extension of spacetime, it is irrelevant for our shock
results, and even for the general location of the (would be) $r=0$
spacelike singularity. We demonstrate this by a simple calculation
of Planck length in our units ($c=G=M_{0}=1$). From units considerations,
the reduced Planck constant satisfy $\hbar\propto GM_{0}^{2}/c$,
so we can find that in our units $\hbar\approx4.74\times10^{-78}$.
\footnote{The calculation requires an assumption regarding the actual value
of the initial black hole mass, $M_{0}$. We take it to be of order
of a stellar BH, $M_{0}=5M_{\odot}\approx10^{31}\, Kg$. Since the
value of Planck length in our units turns out to be inverse proportional
to the mass, our argument is even stronger for supermassive BH, of
mass order of $\sim10^{5}\, M_{\odot}$ or more. %
} Planck length turns out to be a square root of this number in our
units since $l_{P}=\sqrt{\frac{\hbar G}{c^{3}}}\approx2.18\times10^{-39}$.
This number is many orders of magnitude below the lowest $r$ value
in which we maintain numerical reliability (of order $r\sim10^{-2}$). 

We classify our numerical results into three categories of numerical
reliability. We consider the shock detection results to be highly
reliable: (i) they are confirmed by two different and independent
mechanisms, results on timelike and null geodesics; (ii) they begin
quite far from the $r=0$ singularity ($r\approx0.48$ in the case
of two null fluids perturbation, $r\approx0.34$ in the mixed case);
(iii) they are supported by numerical convergence indicators (e.g.
the overlap of $N=320$ and $N=640$ timelike geodesics results in
Figs. \ref{fig:RNV-NoShock}(a), \ref{fig:DNF-shock-2d}(a) and \ref{fig:Mixed-shock-2d}(a)).
The second category contains results we consider reliable: the detection
of a spacelike $r=0$ singularity in the case of two null fluids and
our sharpening rate law analysis. The exact location of the spacelike
singularity might be slightly distorted due to numerical performance
(or resolution limits) issues, but its spacelike nature is a clear
and consistent feature across the grid. Our sharpening rate analysis
is based on highly reliable timelike geodesics data but involves an
interpolation mechanism which may slightly distort the curves shape,
but not alter the basic match between the curves of $\kappa_{-}$
and the derivatives of $-\ln(\Delta\tau)$ (also, since both $\kappa_{-}$
and the derivative curves are interpolated in the same fashion, they
are unlikely to be distorted in different directions). As we have
already elaborated, the detection of a null $r=0$ section of the
singularity in the case of two null fluids belongs to the third category
--- it is uncertain and requires further research. 

Our research could be extended in many directions. Due to the limited
scope of this paper, we have focused our attention on the gravitational
shock (or the shock in the metric function $r$). Nevertheless, the
shock is expected to manifest in the scalar field as well. It could
be interesting to analyze the effect of the null fluid stream on the
shock in the scalar field $\Phi$ in the mixed case, as a ``toy model''
for interaction between two different types of perturbations in the
context of the shock; astrophysical BHs are expected to accrete matter
as well as radiation. It could also be enlightening to study how the
shock manifests, if at all, in various curvature scalars; MO and EO
have not discussed their behaviour specifically in the context of
the shock, but equivalent studies have been made regarding CH, a well
known curvature singularity in the perturbed charged (or spinning)
case (see e.g. Ref. \cite{20-Burko-CScalars-Kerr} for a recent numerical
study in Kerr).

We believe that the most acute next step in the numerical study of
the shock is the extension of the research to spinning BHs. Astrophysical
BHs are expected to be spinning; for instance, in recently detected
gravitational waves events GW150914 \cite{GW150914} and GW151226,
\cite{GW151226} the outcome of the mergers was BHs with a significant
spin, $a/m\sim0.7$. As far as we are aware, there had been no numerical
verification of the shock existence in the spinning case. This numerical
study would be extremely challenging due to the lack of spherical
symmetry. 

A different research direction stemming from our research is the numerical
study of the inner structure of charged BHs perturbed by two null
fluids. Possible concrete objectives are determining the exact nature
of the spacelike $r=0$ singularity and the possible existence of
a null, non-naked $r=0$ singularities. This research would require
an improvement in the numerical performance of our algorithm near
the singularity; both the ODE solver of the constraint equation on
the initial ray $u=u_{0}$ (Eq. (\ref{eq:  ruu})) and the grid bulk
solver of the PDEs need to be ``immunized'' against the singularity.
A possible mechanism may be a special gauge selection. EO have used
a special gauge variant of the maximal-$\sigma$ gauge, called the
singularity approach gauge. This variant was intended to resolve the
approach to a contracting CH on the grid final ray $v=v_{max}$. If
this gauge could be generalized to resolve any approach to $r=0$
(on any $v$ value), it should be a suitable candidate for this research.

\section*{Acknowledgments}

I am indebted to Prof. Amos Ori for many invaluable discussions and
helpful comments. I would also like to thank Prof. Eric Poisson for
his helpful answer regarding the case of two null fluids. This research
was supported by the Israel Science Foundation (Grant No. 1346/07). 

\appendix

\section{Appendix: RNV advanced time coordinate \label{sec:V-RNV} }

\subsection{Definition, numerical calculation and interpretation\label{sub:v-Vaidya-def}}

The definition and calculation of RNV advanced time coordinate is
identical for all the scenarios in this paper, although the meaning
and the interpretation of the coordinate differ. We now describe spacetime
with a new metric, the ingoing RNV metric. The line element is defined
(in the coordinates $(V,r,\theta\varphi))$ as

\begin{equation}
ds^{2}=-(1-\frac{2m(V)}{r}+\frac{Q^{2}}{r^{2}})dV^{2}+2dVdr+r^{2}d\Omega^{2}\,.\label{eq:RNV-metric}
\end{equation}

In order to find RNV advanced time coordinate $V$ as a (numerical)
function of the numerical coordinate $v$, we have applied two different
methods: (i) we calculated $V$ using the coordinate transformation
from RNV metric to our metric (given by the line element of Eq. (\ref{eq:line-element}));
(ii) we considered an outgoing null ray that satisfies $ds^{2}=0$
and $d\Omega^{2}=0$. The results of these two independent calculations
are denoted below as $V_{a}(v)$ and $V_{b}(v)$, accordingly. We
hereby explain both methods and demonstrate that they yield consistent
results.

The standard metric coordinate transformation satisfies $g'_{\mu\nu}=g_{\alpha\beta}\frac{\partial x^{\alpha}}{\partial x^{\mu}}\frac{\partial x^{\beta}}{\partial x^{\nu}}$.
In particular, since $\frac{\partial V}{\partial u}=0$, $g_{uv}$
yields

\[
g_{uv}=-\frac{e^{\sigma}}{2}=g_{\alpha\beta}\frac{\partial x^{\alpha}}{\partial u}\frac{\partial x^{\beta}}{\partial v}=\frac{\partial r}{\partial u}\frac{\partial V}{\partial v}\,.
\]

\noindent We can isolate $\frac{\partial V}{\partial v}$ and integrate
to obtain

\begin{equation}
V_{a}(v)=-\intop\frac{e^{\sigma}}{2r,_{u}}\, dv\,.\label{eq:Vaidya1Analytical}
\end{equation}

Alternately, we may consider an outgoing null ray $u=const$. Since
this geodesic satisfies $ds^{2}=0$ and $d\Omega^{2}=0$, we obtain
from the line element (Eq. (\ref{eq:RNV-metric}))

\[
2dVdr=(1-\frac{2m(V)}{r}+\frac{Q^{2}}{r^{2}})dV^{2}\,.
\]

\noindent Now we can isolate $\frac{dV}{dr}$ and integrate to obtain

\begin{equation}
V_{b}(v)=\intop\frac{2}{(1-\frac{2m(V)}{r}+\frac{Q^{2}}{r^{2}})}\, dr\,.\label{eq:Vaidya2Analytical}
\end{equation}

\noindent In both methods the integration is performed retroactively,
after the functions $r,_{u}(v),\: r(v)$ and $m(v)$ are known. We
use a simple numerical integration scheme that replaces Eq. (\ref{eq:Vaidya1Analytical})
with

\begin{equation}
V_{a}(v+\Delta v)=V_{a}(v)-\frac{e^{(\sigma(v)+\sigma(v+\Delta v))/2}}{r,_{u}(v)+r,_{u}(v+\Delta v)}\Delta v\,,\label{eq:Vaidya1Numerical}
\end{equation}

\noindent and Eq. (\ref{eq:Vaidya2Analytical}) with

\begin{equation}
V_{b}(v+\Delta v)=V_{b}(v)+\frac{4\Delta r}{2-\frac{2m(v)}{r(v)}-\frac{2m(v+\Delta v)}{r(v+\Delta v)}+\frac{Q^{2}}{r(v)^{2}}+\frac{Q^{2}}{r(v+\Delta v)^{2}}}\,.\label{eq:Vaidya2Numerical}
\end{equation}

\noindent This integration scheme is second order accurate. We fix
the integration constant, or the origin of the coordinate, by the
choice to identify $V=0$ with the end of the scalar-field pulse in
the mixed case $(v=7)$. We opted to keep this choice in the single
null fluid case and the two null fluids case as well, despite the
absence of the scalar-field pulse, for the sake of uniformity. 

In principle, the procedure implied by Eq. (\ref{eq:Vaidya1Numerical})
and (\ref{eq:Vaidya2Numerical}) could be performed along any outgoing
null ray and yield different variants of RNV advanced time coordinate.
We have calculated two variants: along the initial ray $u=u_{0}$
and along the EH, $u=u_{h}$. We call the first variant the initial
ray RNV advanced time coordinate, denoted $V$, and the second variant
the horizon RNV advanced time coordinate, denoted $\widetilde{V}$.
We employed both methods $a$ and $b$ to calculate $V$ (although
we display in the rest of the paper $V_{a}$ as $V$), and we employed
the second method to calculate $\widetilde{V}$ (since $r,_{u}$ along
the horizon was more difficult to obtain). 

While the derivation of the coordinate according to Eq. (\ref{eq:Vaidya1Numerical})
and (\ref{eq:Vaidya2Numerical}) is rather straightforward, the interpretation
of the coordinate is more elusive. For instance, in the presence of
outgoing null fluid or a (strong) self-gravitating scalar field, the
mass function $m$ is no longer a function of $v$ alone. The derivation
of RNV advanced time on such patch holds little meaning --- the coordinate
functions as a RNV coordinate along a single null ray. A similar claim
holds for the usage of RNV advanced time derived in one (eligible)
spacetime patch in another (eligible) spacetime patch, separated from
the first patch by a region of outgoing null fluid or scalar field
perturbations. The mass function in both patches may be a function
of $v$ alone, but it is a different function of $v$. Thus a coordinate
calculated in the first patch does not function as RNV advanced time
coordinate in the other patch --- it is disconnected from the mass
function of the second patch.

In the scenarios considered in this paper, the situation is relatively
simple to analyze. In the single null fluid case, RNV advanced time
is an ``exact'' RNV coordinate on the entire domain of integration,
since the mass function $m$ is indeed a function of $v$ alone on
the whole grid. In the case of two null fluids, $V$ is an exact RNV
coordinate up to the beginning of the outgoing null fluid region at
$u=u_{2}$ (i.e., $V$ functions as an exact RNV coordinate on the
region $u_{0}\leq u\leq u_{2}$). For $u>u_{2}$, $V$ is just an
approximated RNV coordinate. $\widetilde{V}$ is an approximated RNV
coordinate on the whole grid (since the horizon is located inside
the outgoing null fluid region). Still, since the effect of the outgoing
null fluid is relatively minor up to the innermost part of the BH,
this approximation is sensible; one might argue that deep inside the
BH $\widetilde{V}$ is a ``better'' approximated RNV coordinate,
since the extent of outgoing null fluid that separates between the
horizon and this region is smaller. In the mixed case, both $V$ and
$\widetilde{V}$ are approximated RNV coordinates on the whole grid
due to the presence of a self-gravitating scalar field; however, at
late times, when the scalar field scatters and decays, the quality
of the approximation improves, and asymptotically both coordinates
functions as a ``good'' RNV coordinate. This type of behavior may
be termed as the behavior of a RNV-like advanced time coordinate,
which is the equivalent of Eddington-like advanced time coordinate
defined by EO.

Fig. \ref{fig:V-Vaidya} displays numerical results for $V_{a}(v),V_{b}(v)$
and $\widetilde{V}(v)$ in the case of two null fluids (Fig. \ref{fig:V-Vaidya}(a))
and in the mixed case (Fig. \ref{fig:V-Vaidya}(b)). The single null
fluid case was omitted as it is less interesting in this context ($V_{a}$
and $V_{b}$ are identical to those of the two null fluids case and
$\widetilde{V}$ is not needed). As expected, the differences between
$V_{a}$ and $V_{b}$ are negligibly small. The difference between
$V(v)$ and $\widetilde{V}(v)$ is significant in the case of two
null fluids and smaller in the mixed case. 

\begin{center}
\begin{figure}[H]
\begin{centering}
\includegraphics[scale=0.35]{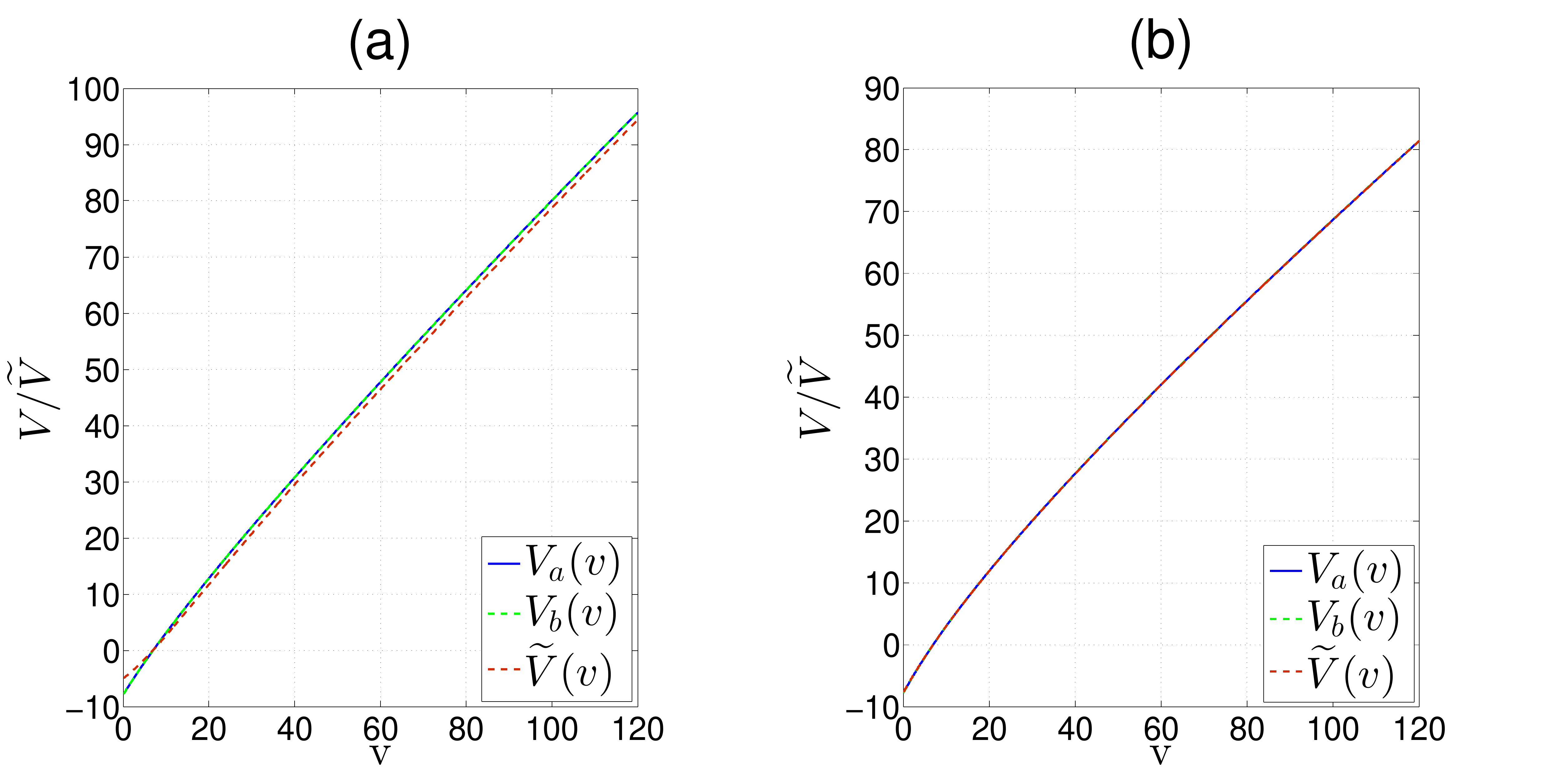}
\par\end{centering}

\protect\caption{\label{fig:V-Vaidya} Numerical results for RNV advanced time coordinate
as a function of the numerical $v$ coordinate. Panel (a) describes
the case of two null fluids and panel (b) describes the mixed case.
The initial ray RNV advanced time coordinate $V$ is evaluated in
two methods, $V_{a}$ and $V_{b}$, and the horizon RNV advanced time
coordinate $\widetilde{V}$ is evaluated by a single method ($b$).
While the differences between $V_{a}$ and $V_{b}$ are extremely
small in both cases (roundoff oriented, $\lesssim10^{-14}$ in the
case of two null fluids and $\lesssim10^{-12}$ in the mixed case)
the difference between $\widetilde{V}$ and $V$ is noticeable in
the case of two null fluids (maximal difference is $|V-\widetilde{V}|\approx2.9$)
and small in the mixed case (maximal difference is $|V-\widetilde{V}|\approx0.06$,
although all three curves overlap in the scale of panel (b)). }
\end{figure}

\par\end{center}

\subsection{Linear mass function derivation\label{sub:Linear-mass-calc}}

We turn next to consider the specific ingoing null fluid stream defined
by Eq. (\ref{eq:Tvv_shape}); we demonstrate that it has a linear
contribution to the mass function $m(V)$ at late times. This property
is gauge dependent and may vanish in a gauge transformation. We begin
with a previous result of Poisson and Israel, \cite{7-Poisson-Israel-mass-function}

\[
m,_{a}=4\pi r^{2}T_{a}^{\:\, b}r,_{b}\,,
\]

\noindent where $T_{ab}\equiv T_{\mu\nu}^{NF}$ is the energy-momentum
tensor of the null fluid. We discuss here the ingoing null fluid stream
on the initial ray $u=u_{0}$, so we are interested in finding $m,_{v}$.
After the appropriate index lowering and submitting $g^{uv}=2e^{-\sigma}$,
we obtain

\[
m,_{v}=4\pi r^{2}g^{uv}T_{vv}^{NF}r,_{u}=-8\pi r^{2}e^{-\sigma}T_{vv}^{NF}r,_{u}\,.
\]
Inserting the nonvanishing part of $T_{vv}^{NF}(u_{0},v)$ from Eq.
(\ref{eq:Tvv_shape}) yields

\[
m,_{v}(u_{0},v)=-8\pi A_{1}e^{-\sigma}(1-e^{r_{0}-r})^{2}\frac{1}{r,_{u}}\,.
\]

\noindent We are interested in finding $m,_{V}=m,_{v}\frac{dv}{dV}$.
We can see from Eq. (\ref{eq:Vaidya1Analytical}) that $\frac{dv}{dV}=-2r,_{u}e^{-\sigma}$,
so

\[
m,_{V}(u_{0},V)=16\pi A_{1}e^{-2\sigma}(1-e^{r_{0}-r})^{2}\,.
\]

\noindent Recalling that on our initial ray $\sigma(u_{0},v)=0$,
we get the final result

\begin{equation}
m,_{V}(u_{0},V)=16\pi A_{1}(1-e^{r_{0}-r})^{2}\,.
\end{equation}

\noindent The factor $(1-e^{r_{0}-r})^{2}$ approaches $1$ with the
rise in $r$, yielding the required constant $m,_{V}$ (or linear
$m(V)$). Fig. \ref{fig:m_V_ofV} displays the numerical $m,_{V}(V)$
and confirms that it fits the expected behavior.

\begin{center}
\begin{figure}[H]
\begin{centering}
\includegraphics[scale=0.35]{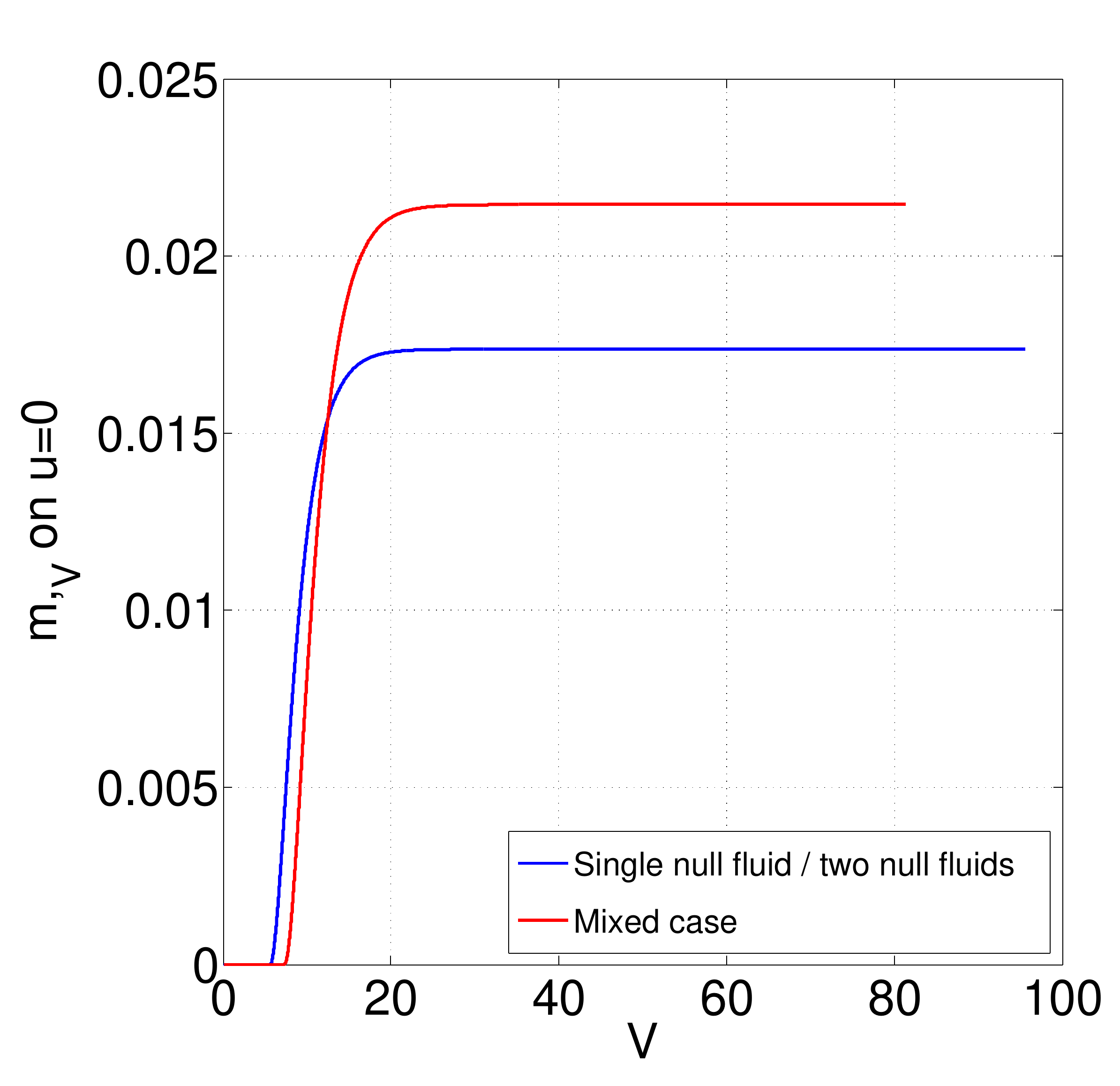}
\par\end{centering}

\protect\caption{\label{fig:m_V_ofV} Numerical results for $m,_{V}(V)$ on the initial
ray $u_{0}=0$. The blue curve fits the single null fluid and two
null fluids scenarios, which share the same ingoing null fluid stream,
and the red curve fits the mixed case. In both cases $m,_{V}$ rises
smoothly from zero and approaches a constant value, implying a linear
mass function $m(V)$ at late times $(V\gtrsim30)$.}
\end{figure}

\par\end{center}

\end{document}